\documentclass[aps,prd,groupedaddress,showpacs,nofootinbib]{revtex4-1}

\usepackage[hmargin=2.54cm,vmargin=2.54cm]{geometry}   
\usepackage{amsmath}
\usepackage{multirow}
\usepackage{wrapfig}
\usepackage{mathrsfs}
 \usepackage{amsthm}
\usepackage{graphicx}

\begin{document}


\title{Calculation of High Energy Neutrino-Nucleon Cross Sections and Uncertainties
Using the MSTW Parton Distribution Functions and Implications for Future Experiments}


\author{Amy~Connolly\footnote{Prior to September 2010, at University College London.}}
\email{connolly@mps.ohio-state.edu}
\affiliation{Department of Physics and CCAPP, The Ohio State University, 191 West Woodruff Avenue, Columbus, Ohio  43210}
\author{Robert~S.~Thorne}
\author{David~Waters}
\affiliation{Department of Physics and Astronomy, University College London, Gower Street, London WC1E 6BT}

\date{\today}



\begin{abstract}
We present a new calculation of the
cross sections for charged current (CC) and neutral current (NC) 
$\nu N$ and $\bar{\nu} N$ interactions in the neutrino 
energy range $10^{4}<E_{\nu}<10^{12}$~GeV
using the most recent MSTW parton distribution functions (PDFs), 
MSTW~2008.  We also present the associated uncertainties
propagated from the PDFs, as well as parametrizations of 
the cross section central values, their uncertainty bounds, 
and the inelasticity distributions for ease of use in Monte Carlo simulations.  
For the latter we only provide parametrizations for energies above $10^7$~GeV. 
Finally, we assess the feasibility of future neutrino 
experiments to constrain the $\nu N$ cross section in the
ultra-high energy (UHE) regime using a technique that is independent
of the flux spectrum of incident neutrinos. 
A significant deviation from the predicted Standard Model cross sections
could be an indication of new physics, such as extra space-time
dimensions, and we present expected constraints on such models
as a function of the number of events observed in 
a future subterranean neutrino
detector.
\end{abstract}



\maketitle


\section{Introduction}
\label{intro}

Neutrino experiments are closing in on neutrinos in the ultra-high energy (UHE) 
regime, where a diffuse neutrino flux, first predicted by Berezinsky and 
Zatsepin~\cite{Berezinsky:1969zz,Berezinsky:1970}, is expected to result from interactions between UHE cosmic rays and cosmic 
microwave background photons through what is known as the 
Greisen, Zatsepin and Kuzmin 
(GZK) process~\cite{Greisen:1966jv,Zatsepin:1966jv}.  
Neutrinos in this energy regime probe higher center-of-mass (COM) energies 
than those accessible by human-made accelerators through their interactions in the earth.  
For example, the COM energy of a $10^{9}$~GeV neutrino incident on a nucleon 
at rest
is 45~TeV.  Thus, a measurement of neutrino-nucleon ($\nu N$) 
 cross sections in the UHE
regime could be sensitive to either new physics scenarios 
such as extra space-time dimensions or unexpected behavior of parton distribution 
functions (PDFs) at Bjorken-$x$ smaller than
that accessible by current experiments (from here on, any $\nu$ refers to
both neutrino and anti-neutrino unless otherwise stated)~\cite{Anchordoqui:2006ta}.  
However, before the significance 
of any $\nu N$ cross section measurement can be assessed, the
uncertainties on the Standard Model (SM) expectation 
must be quantified based on the diverse body of current experimental 
constraints.  

The paper is composed of two parts.  In Section~\ref{part:calc}, we perform
a new calculation of $\nu N$ cross sections and their associated PDF uncertainties 
for neutrino energies $E_{\nu}>10^4$~GeV using the MSTW~2008 PDF set.
In Section~\ref{sec:nuNcs}, we review the expressions for the $\nu N$ cross sections
in terms of the quark PDFs.
In Section~\ref{sec:results} we present the results of our  
cross section calculations, associated uncertainties and their 
energy dependent parametrizations. Next, in Section~\ref{sec:dsigmadydx}
we discuss the differential cross sections, and
parametrize the inelasticity 
distributions in an energy
dependent way. We also show a few
select distributions in $x$.  In Section~\ref{sec:correlations},
we calculate the correlations between the uncertainties across
energies.

In Section~\ref{part:meas}, we propose to constrain the
UHE $\nu N$ cross section in future sub-terranean neutrino experiments
using a technique that is independent of the incident
flux spectrum of neutrinos and present projected
constraints on models with enhanced cross sections due to extra
space-time dimensions.

\section{Neutrino-Nucleon Cross Section}
\label{part:calc}

\subsection{Methodology}
\label{sec:nuNcs}

The $\nu N$ cross section for charged current (CC) interactions on an isoscalar target is given 
by~\footnote{Natural units, $\hbar=c=1$, are assumed throughout}: 
\begin{equation}
\sigma_{CC}\left(E_{\nu}\right)=\frac{2G_F^2M_NE_{\nu}}{\pi} \int_{0}^1\int_{0}^1 dy~dx \left(\frac{M_W^2}{Q^2+M_W^2}\right)^2 \left[q+\left(1-y\right)^2 \bar{q}\right]
\end{equation}
with quark and antiquark densities given by $q=(d+u)/2+s+b$ and $\bar{q}=(\bar{d}+\bar{u})/2+c+t$.  
In all of the equations in this paper we assume that 
a quark distribution function is equivalent to the corresponding antiparticle distribution 
except for $u$ and $d$.  
In MSTW~2008, $t=0$ and it is not strictly true that $s=\bar{s}$, but these are negligible effects 
for our calculations.  Here, $G_F=1.17\times10^{-5}$~GeV$^{-2}$ is the Fermi coupling constant and 
$M_N$ is the nucleon mass for which we use the proton mass, 0.938~GeV. The mass of the $W$ boson 
$M_W=80.398$~GeV, $E_{\nu}$ is the incident neutrino energy, and $x$ and $y$ are the parton 
momentum fraction (Bjorken-$x$) and the inelasticity, respectively.  
\

Likewise, the neutral current (NC) $\nu N$ cross section is given by:
\begin{equation}
\sigma_{NC}\left(E_{\nu}\right)=\frac{2G_F^2M_NE_{\nu}}{\pi}\int_{0}^1\int_{0}^1  dy~dx
\left(\frac{M_Z^2}{Q^2+M_Z^2}\right)^2 \left[q^0+\left(1-y\right)^2 \bar{q^0}\right]   
\end{equation}
where $M_Z$ is the $Z$ mass.  Then 
\begin{align}
q^0&=\frac{u+d}{2}\left(L_u^2+L_d^2 \right)+
\frac{\bar{u}+\bar{d}}{2}\left(R_u^2+R_d^2\right)+\\
&\left(s+b\right)\left(L_d^2+R_d^2\right)+
\left(c+t\right)\left(L_u^2+R_u^2\right)
\end{align}
and
\begin{align}
\bar{q^0}=&\frac{u+d}{2}\left(R_u^2+R_d^2 \right)+
\frac{\bar{u}+\bar{d}}{2}\left(L_u^2+L_d^2\right)\\
&+\left(s+b\right)\left(L_d^2+R_d^2\right)+
\left(c+t\right)\left(L_u^2+R_u^2\right)
\end{align}
with $L_u=1-4/3\cdot x_W$, $L_d=-1+2/3\cdot x_W$, $R_u=-4/3\cdot x_W$ and 
$R_d=2/3\cdot x_W$ where $x_W=\sin^2{\theta_W}=0.226$. 
For the $\bar{\nu}$N cross sections, the above
equations are the same with 
each quark distribution function replaced with the
corresponding antiparticle distribution and vice versa,
so that $q \leftrightarrow \bar{q}$, 
$q^0 \leftrightarrow \bar{q^0}$.

We use the parton distribution functions calculated by A.D.~Martin~{\sl et~al.}
known as ``MSTW~2008''~\cite{Martin:2009iq}. These PDFs are the latest update to a 
series that began with the MRS PDFs twenty years ago, which were the first
global next-to-leading-order (NLO) PDF analysis.  The MSTW~2008 set 
incorporates improvements in the precision and kinematic range of recent
measurements as well as improved theoretical developments which
make the global analysis more reliable.  
The publication of the
MSTW~2008 set was particularly 
timely in view of the start of data taking at the Large Hadron Collider (LHC).

\subsection{Cross Sections}
\label{sec:results}

\begin{figure} 
\includegraphics[width=3.4in]{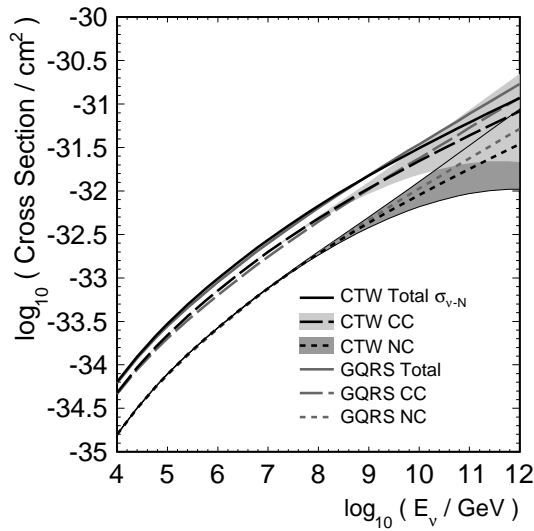} 
   \caption{Calculated $\nu N$ CC and NC cross sections.  In this plot we compare this work (CTW), 
    with the shaded bands representing the associated uncertainties due to PDFs, to those in 
    Gandhi {\it et al.} (GQRS).  Thin black lines bound the 
NC uncertainties so that they remain visible where they overlap with the CC bounds.  \label{fig:cs}}
\end{figure}

\begin{figure}
   \includegraphics[width=3.4in]{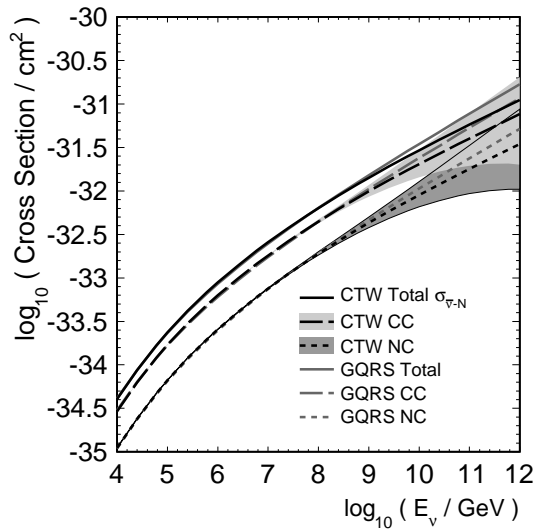}
    \caption{Calculated $\bar{\nu} N$ CC and NC cross sections. In this plot we compare this 
    work (CTW), with the shaded bands representing the associated uncertainties due to PDFs, 
    to those in GQRS.\label{fig:cs_nubar}}
\end{figure}

Figs.~\ref{fig:cs} and~\ref{fig:cs_nubar}
shows the results of our $\nu N$ and $\bar{\nu}N$
cross section calculations compared to the previous calculations by 
Gandhi {\it et al.} (GQRS) \cite{Gandhi:1998ri}.  
These results are summarized numerically 
in Tables~\ref{tab:nuN} and~\ref{tab:nubarN}.
With regard to uncertainties, the latter paper only
states that they find the uncertainties in the $\nu N$ cross sections
to be at most a factor of $2^{\pm1}$. Recently Cooper-Sarkar and Sarkar 
(CSS) \cite{CooperSarkar:2007cv}
also published CC $\nu N$ cross sections for energies in the range 
$100<s<10^{12}$~GeV$^2$ where $\sqrt{s}$ is the COM energy of the interaction.
There has also been a recent investigation into the dependence on the 
number of active heavy quarks in ~\cite{Jeong:2010za}.
We find good agreement with the central values of both the GQRS and CSS 
calculations within our uncertainties.
We note that the explicit evolution of the MSTW 2008 PDFs only
takes place down to the lowest value of the grid points, i.e. $x=10^{-6}$. 
Below this the values for the 
central set and each eigenvector for the error sets are extrapolated   
linearly in $\ln(1/x)$. Within the region of the grids the accuracy of the 
NLO and NNLO evolution has been checked to small fractions of a percent, 
see Section 3 of \cite{Martin:2009iq}. 
\footnote{A numerical inaccuracy in the LO 
evolution at very small $x$, albeit very much less than the uncertainty,
has been pointed out in \cite{Block:2010fk}. This is unique to LO
due to the extreme singular behaviour of the small-$x$ gluon in this case.  
It will be corrected in future sets.}


\begin{figure}
  \includegraphics[width=3.4in]{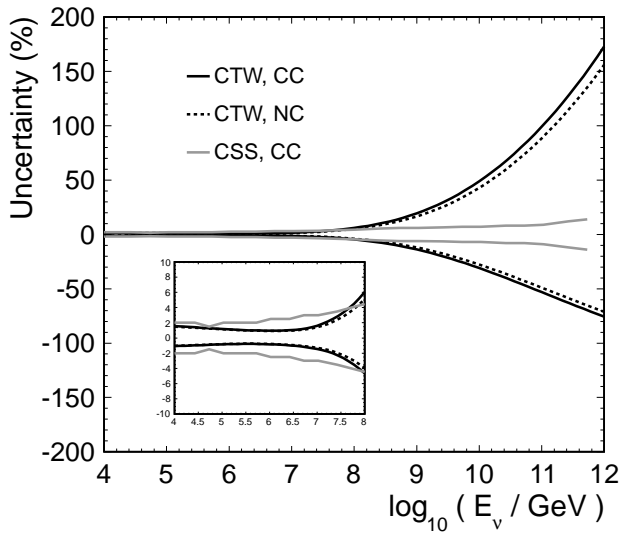}
\caption{\label{fig:uncertainties} Uncertainties on the calculated cross sections due to PDFs.  
We compare this work (CTW) to CSS.}
\end{figure}

Fig.~\ref{fig:uncertainties} compares the uncertainties on the cross section calculations 
for the range of energies being considered. The uncertainties on our calculations 
are dramatically different from those reported by CSS for $E_{\nu}\gtrsim 10^8$~GeV.  
The difference is due to a different parametrization of the gluon
parton distribution $g\left(x\right)$.  The CSS fit to the HERA data allows
a very good fit with the gluon distribution having an 
$x$ dependence of the form $g\left(x\right)~\propto~x^{\delta}$.
However, MSTW~2008 finds that a sum of two terms with different powers
$xg\left(x\right)~\propto~A_1 x^{\delta_1}+A_2 x^{\delta_2}$ gives a 
better fit to the global data set. This is partially due to the global fit
requiring a slightly larger value of the strong coupling $\alpha_S$ and 
consequently less gluon to drive small-$x$ structure function evolution. It
is also found that Tevatron jet data prefer a larger high-$x$ gluon distribution, hence 
allowing a smaller gluon distribution at  small-$x$ 
from the momentum sum rule. 
(It is shown in \cite{Martin:2010db} that fitting the newer combined
HERA data in \cite{:2009wt} results in no very 
significant change to the MSTW PDFs.) 

As well as producing the best fit, the greater flexibility brought about 
by including two terms in the parametrization results in more rapid
expansion of the allowed range in $g\left(x\right)$ 
at low-$x$ beyond the reach of current experiments, as illustrated in 
Fig. 16 of \cite{Martin:2009iq}. With the parametrization used by CSS, the 
uncertainty can only grow as a 
function of $\ln\left(1/x\right)$ in this region as discussed in Section 6.5 
of \cite{Martin:2009iq}.
We notice that the point at which our uncertainty starts to exceed that of CSS
to a significant extent is indeed when the constraint due to HERA data is 
starting to disappear, i.e. when the dominant $x$ values contributing to the 
cross section are $x=10^{-5}$ or lower. At very high neutrino energy when 
the $x$ 
values probed are typically well below $x=10^{-5}$ our uncertainty on the 
cross section becomes very large. Hence, we conclude that a good
measurement of the cross section within this range will provide the first
direct constraint on the extremely small-$x$ PDFs, and can give us the first
true indication of their central value as well as reducing the uncertainty 
significantly.

At the lower end of the uncertainty bands the cross sections do contain contributions from 
PDFs that have become negative. It is difficult to know whether this
is really a problem. At low orders in $\alpha_S$ and leading twist 
perturbation theory this could lead to negative cross sections. However, we 
have the possibility of both large $\ln(1/x)$ perturbative corrections and 
higher-twist nonlinear effects in this regime which could alter this 
conclusion. Hence, our uncertainty at the very highest energies can be 
thought of as acknowledging the possibility of theoretical corrections in 
this regime.  

The recent comparative study of high energy neutrino 
cross sections in \cite{Goncalves:2010ay} illustrates the variation between 
models at very high energies due to theoretical assumptions. 
(Note the variation at lower 
energies between models in Figs.~\ref{fig:cs} and~\ref{fig:cs_nubar} 
of this paper is likely due to the 
omission of terms required at higher $x$ \cite{Thorne:2005kj}).

\begin{table}
    \caption{Cross Sections for $\nu$N.\label{tab:nuN}}
   \begin{center}
      \begin{tabular}{p{2cm}p{2cm}p{2cm}p{2cm}} \hline \hline
$E_{\nu}$ (GeV) & $\sigma_{\rm{CC}}$ (cm$^{-2}$) & $\sigma_{\rm{NC}}$ (cm$^{-2}$) & $\sigma_{\rm{tot}}$ (cm$^{-2}$) \\ \hline
 1 $\times 10^{4}$ 	 &	0.48$\times 10^{-34}$ 	&	0.16$\times 10^{-34}$ 	&	0.63$\times 10^{-34}$ \\ 
 2.5 $\times 10^{4}$ 	 &	0.93$\times 10^{-34}$ 	&	0.32$\times 10^{-34}$ 	&	0.12$\times 10^{-33}$ \\ 
 6 $\times 10^{4}$ 	 &	0.16$\times 10^{-33}$ 	&	0.57$\times 10^{-34}$ 	&	0.22$\times 10^{-33}$ \\ 
 1 $\times 10^{5}$ 	 &	0.22$\times 10^{-33}$ 	&	0.78$\times 10^{-34}$ 	&	0.3$\times 10^{-33}$ \\ 
 2.5 $\times 10^{5}$ 	 &	0.36$\times 10^{-33}$ 	&	0.13$\times 10^{-33}$ 	&	0.49$\times 10^{-33}$ \\ 
 6 $\times 10^{5}$ 	 &	0.56$\times 10^{-33}$ 	&	0.21$\times 10^{-33}$ 	&	0.77$\times 10^{-33}$ \\ 
 1 $\times 10^{6}$ 	 &	0.72$\times 10^{-33}$ 	&	0.27$\times 10^{-33}$ 	&	0.98$\times 10^{-33}$ \\ 
 2.5 $\times 10^{6}$ 	 &	0.11$\times 10^{-32}$ 	&	0.41$\times 10^{-33}$ 	&	0.15$\times 10^{-32}$ \\ 
 6 $\times 10^{6}$ 	 &	0.16$\times 10^{-32}$ 	&	0.61$\times 10^{-33}$ 	&	0.22$\times 10^{-32}$ \\ 
 1 $\times 10^{7}$ 	 &	0.2$\times 10^{-32}$ 	&	0.76$\times 10^{-33}$ 	&	0.27$\times 10^{-32}$ \\ 
 2.5 $\times 10^{7}$ 	 &	0.29$\times 10^{-32}$ 	&	0.11$\times 10^{-32}$ 	&	0.4$\times 10^{-32}$ \\ 
 6 $\times 10^{7}$ 	 &	0.4$\times 10^{-32}$ 	&	0.16$\times 10^{-32}$ 	&	0.56$\times 10^{-32}$ \\ 
 1 $\times 10^{8}$ 	 &	0.48$\times 10^{-32}$ 	&	0.19$\times 10^{-32}$ 	&	0.67$\times 10^{-32}$ \\ 
 2.5 $\times 10^{8}$ 	 &	0.67$\times 10^{-32}$ 	&	0.27$\times 10^{-32}$ 	&	0.94$\times 10^{-32}$ \\ 
 6 $\times 10^{8}$ 	 &	0.91$\times 10^{-32}$ 	&	0.36$\times 10^{-32}$ 	&	0.13$\times 10^{-31}$ \\ 
 1 $\times 10^{9}$ 	 &	0.11$\times 10^{-31}$ 	&	0.43$\times 10^{-32}$ 	&	0.15$\times 10^{-31}$ \\ 
 2.5 $\times 10^{9}$ 	 &	0.14$\times 10^{-31}$ 	&	0.58$\times 10^{-32}$ 	&	0.2$\times 10^{-31}$ \\ 
 6 $\times 10^{9}$ 	 &	0.19$\times 10^{-31}$ 	&	0.77$\times 10^{-32}$ 	&	0.27$\times 10^{-31}$ \\ 
 1 $\times 10^{10}$ 	 &	0.22$\times 10^{-31}$ 	&	0.9$\times 10^{-32}$ 	&	0.31$\times 10^{-31}$ \\ 
 2.5 $\times 10^{10}$ 	 &	0.29$\times 10^{-31}$ 	&	0.12$\times 10^{-31}$ 	&	0.41$\times 10^{-31}$ \\ 
 6 $\times 10^{10}$ 	 &	0.37$\times 10^{-31}$ 	&	0.15$\times 10^{-31}$ 	&	0.53$\times 10^{-31}$ \\ 
 1 $\times 10^{11}$ 	 &	0.43$\times 10^{-31}$ 	&	0.18$\times 10^{-31}$ 	&	0.61$\times 10^{-31}$ \\ 
 2.5 $\times 10^{11}$ 	 &	0.56$\times 10^{-31}$ 	&	0.23$\times 10^{-31}$ 	&	0.8$\times 10^{-31}$ \\ 
 6 $\times 10^{11}$ 	 &	0.72$\times 10^{-31}$ 	&	0.3$\times 10^{-31}$ 	&	0.1$\times 10^{-30}$ \\ 
 1 $\times 10^{12}$ 	 &	0.83$\times 10^{-31}$ 	&	0.35$\times 10^{-31}$ 	&	0.12$\times 10^{-30}$ \\ \hline \hline
     \end{tabular}
    \end{center}
\end{table}

\begin{table}
    \caption{Cross Sections for $\bar{\nu}$N.\label{tab:nubarN}}
   \begin{center}
      \begin{tabular}{p{2cm}p{2cm}p{2cm}p{2cm}} \hline \hline
$E_{\bar{\nu}}$ (GeV) & $\sigma_{\rm{CC}}$ (cm$^{-2}$) & $\sigma_{\rm{NC}}$ (cm$^{-2}$) & $\sigma_{\rm{tot}}$ (cm$^{-2}$) \\ \hline
 1 $\times 10^{4}$ 	 &	0.29$\times 10^{-34}$ 	&	0.11$\times 10^{-34}$ 	&	0.4$\times 10^{-34}$ \\ 
 2.5 $\times 10^{4}$ 	 &	0.63$\times 10^{-34}$ 	&	0.24$\times 10^{-34}$ 	&	0.87$\times 10^{-34}$ \\ 
 6 $\times 10^{4}$ 	 &	0.12$\times 10^{-33}$ 	&	0.47$\times 10^{-34}$ 	&	0.17$\times 10^{-33}$ \\ 
 1 $\times 10^{5}$ 	 &	0.17$\times 10^{-33}$ 	&	0.67$\times 10^{-34}$ 	&	0.24$\times 10^{-33}$ \\ 
 2.5 $\times 10^{5}$ 	 &	0.3$\times 10^{-33}$ 	&	0.12$\times 10^{-33}$ 	&	0.42$\times 10^{-33}$ \\ 
 6 $\times 10^{5}$ 	 &	0.49$\times 10^{-33}$ 	&	0.2$\times 10^{-33}$ 	&	0.68$\times 10^{-33}$ \\ 
 1 $\times 10^{6}$ 	 &	0.63$\times 10^{-33}$ 	&	0.26$\times 10^{-33}$ 	&	0.89$\times 10^{-33}$ \\ 
 2.5 $\times 10^{6}$ 	 &	0.98$\times 10^{-33}$ 	&	0.4$\times 10^{-33}$ 	&	0.14$\times 10^{-32}$ \\ 
 6 $\times 10^{6}$ 	 &	0.15$\times 10^{-32}$ 	&	0.6$\times 10^{-33}$ 	&	0.21$\times 10^{-32}$ \\ 
 1 $\times 10^{7}$ 	 &	0.18$\times 10^{-32}$ 	&	0.76$\times 10^{-33}$ 	&	0.26$\times 10^{-32}$ \\ 
 2.5 $\times 10^{7}$ 	 &	0.26$\times 10^{-32}$ 	&	0.11$\times 10^{-32}$ 	&	0.37$\times 10^{-32}$ \\ 
 6 $\times 10^{7}$ 	 &	0.37$\times 10^{-32}$ 	&	0.16$\times 10^{-32}$ 	&	0.52$\times 10^{-32}$ \\ 
 1 $\times 10^{8}$ 	 &	0.45$\times 10^{-32}$ 	&	0.19$\times 10^{-32}$ 	&	0.64$\times 10^{-32}$ \\ 
 2.5 $\times 10^{8}$ 	 &	0.62$\times 10^{-32}$ 	&	0.27$\times 10^{-32}$ 	&	0.88$\times 10^{-32}$ \\ 
 6 $\times 10^{8}$ 	 &	0.84$\times 10^{-32}$ 	&	0.36$\times 10^{-32}$ 	&	0.12$\times 10^{-31}$ \\ 
 1 $\times 10^{9}$ 	 &	0.99$\times 10^{-32}$ 	&	0.43$\times 10^{-32}$ 	&	0.14$\times 10^{-31}$ \\ 
 2.5 $\times 10^{9}$ 	 &	0.13$\times 10^{-31}$ 	&	0.58$\times 10^{-32}$ 	&	0.19$\times 10^{-31}$ \\ 
 6 $\times 10^{9}$ 	 &	0.17$\times 10^{-31}$ 	&	0.77$\times 10^{-32}$ 	&	0.25$\times 10^{-31}$ \\ 
 1 $\times 10^{10}$ 	 &	0.2$\times 10^{-31}$ 	&	0.9$\times 10^{-32}$ 	&	0.29$\times 10^{-31}$ \\ 
 2.5 $\times 10^{10}$ 	 &	0.27$\times 10^{-31}$ 	&	0.12$\times 10^{-31}$ 	&	0.39$\times 10^{-31}$ \\ 
 6 $\times 10^{10}$ 	 &	0.35$\times 10^{-31}$ 	&	0.15$\times 10^{-31}$ 	&	0.5$\times 10^{-31}$ \\ 
 1 $\times 10^{11}$ 	 &	0.4$\times 10^{-31}$ 	&	0.18$\times 10^{-31}$ 	&	0.58$\times 10^{-31}$ \\ 
 2.5 $\times 10^{11}$ 	 &	0.52$\times 10^{-31}$ 	&	0.23$\times 10^{-31}$ 	&	0.75$\times 10^{-31}$ \\ 
 6 $\times 10^{11}$ 	 &	0.66$\times 10^{-31}$ 	&	0.3$\times 10^{-31}$ 	&	0.96$\times 10^{-31}$ \\ 
 1 $\times 10^{12}$ 	 &	0.77$\times 10^{-31}$ 	&	0.35$\times 10^{-31}$ 	&	0.11$\times 10^{-30}$ \\ \hline \hline
     \end{tabular}
    \end{center}
\end{table}

\subsubsection{Parametrizations}
For ease of use in Monte Carlo simulations, we have parametrized the cross sections in the 
energy range $4<\varepsilon<12$, where $\varepsilon\equiv\log_{10}(E_{\nu} / \rm{GeV})$,  
with an expression of the following form:
\begin{equation}
\begin{array}{lcc}
\label{eq:parameterize}
 \log_{10}\left[ \sigma\left( \varepsilon \right) / \rm{cm}^2\right]& = & C_1 + C_2 \cdot \ln\left(\varepsilon-C_0\right) \\
&& + C_3 \cdot \ln^2\left(\varepsilon-C_0\right) \\
&& + C_4 / \ln\left(\varepsilon-C_0\right). \\
\end{array}
\end{equation}
\begin{table}
    \caption{Coefficients for parametrizing the cross sections according to Equation~\ref{eq:parameterize}.\label{tab:parameterize}
 }

   \begin{center}
      \begin{tabular}{p{1cm}p{1cm}p{1cm}p{1cm}p{1cm}p{1cm}p{1cm}p{1cm}} \hline \hline 
& & $C_{0}$ & $C_{1}$ & $C_{2}$ & $C_{3}$ & $C_{4}$  \\ \hline 
$\nu$ & NC  & \multirow{2}{*}{-1.826} &  \multirow{2}{*}{-17.31}  & -6.448  & \multirow{2}{*}{1.431} & -18.61  \\
$\nu$ & CC &  &  &  -6.406 &  & -17.91  \\ \hline
$\bar{\nu}$ & NC & \multirow{2}{*}{-1.033} & \multirow{2}{*}{-15.95} & -7.296 & \multirow{2}{*}{1.569} & -18.30 \\
$\bar{\nu}$ & CC &  &  & -7.247 &  & -17.72  \\ \hline \hline
      \end{tabular}
    \end{center}
\end{table}

\begin{table}
    \caption{Coefficients for parametrizing the uncertainty bounds on the cross sections according 
    to Equation~\ref{eq:parameterize}.\label{tab:parameterize_bounds} }
   \begin{center}
      \begin{tabular}{p{1cm}p{1cm}p{1cm}p{1cm}p{1cm}p{1cm}p{1cm}p{1cm}} \hline \hline
& & $C_{0}$ & $C_{1}$ & $C_{2}$ & $C_{3}$ & $C_{4}$  \\ \hline 
\multicolumn{7}{c}{upper} \\ \hline
$\nu$ & NC  & \multirow{2}{*}{-1.456} &  32.23  & -32.32  & 5.881 & \multirow{2}{*}{-49.41}  \\
$\nu$ & CC &  & 33.47 & -33.02 & 6.026 &   \\ \hline
$\bar{\nu}$ & NC & \multirow{2}{*}{-2.945} & 143.2 & -76.70 & 11.75 & \multirow{2}{*}{-142.8} \\
$\bar{\nu}$ & CC &  & 144.5 & -77.44  & 11.90 &   \\ \hline \hline
\multicolumn{7}{c}{lower} \\ \hline
$\nu$ & NC  & \multirow{2}{*}{-15.35} &  16.16  & 37.71  & -8.801 & \multirow{2}{*}{-253.1}  \\
$\nu$ & CC &  & 13.86  & 39.84 & -9.205 &  \\ \hline
$\bar{\nu}$ & NC & \multirow{2}{*}{-13.08} & 15.17 & 31.19 & -7.757 & \multirow{2}{*}{-216.1} \\
$\bar{\nu}$ & CC &  & 12.48 & 33.52 & -8.191 &  \\ \hline \hline
      \end{tabular}
    \end{center}
\end{table}

Table~\ref{tab:parameterize} shows the values of the constants for
each of $\nu N$ and $\bar{\nu} N$ interactions, 
CC and NC.  The parametrized cross sections are within approximately 1\% (2\%) of the 
calculated cross sections in the stated energy range for $\nu N$ ($\bar{\nu} N$). 
In Table~\ref{tab:parameterize_bounds} we show the same constants for parametrizing
the upper and lower bounds on the cross sections due to the uncertainties 
derived in this paper.  For the upper bounds, the parametrizations are good to 
approximately 5\% (10\%) for describing our $\nu N$ ($\bar{\nu} N$) calculations.  
For the lower bounds, the parametrizations are at most
approximately 8\% from our $\nu N$ ($\bar{\nu} N$) calculations
until $10^{11.5}<E_{\nu}<10^{12}$~GeV where they deviate by nearly 20\%.

Note that the highest 
power of $\log_{10} E_{\nu}$ required to describe the cross section is 
quadratic, the same as the quadratic 
dependence of the Froissart bound \cite{Froissart:1961ux} on hadron-hadron
cross sections. This shows that although in principle the 
PDFs and cross sections grow quicker than any power of $\log_{10} E_{\nu}$ 
as $E_{\nu} \to \infty$ without some non-linear evolution effects slowing 
the evolution at very small $x$, in practice this has not clearly 
manifested itself in the region of energy we consider. The desire to have a 
parametrization for structure functions manifestly 
consistent with the Froissart bound at all energies has led 
to the results in \cite{Berger:2007ic,Block:2010ud}, which gives rather 
lower predictions than our central values. However, even the upper band of 
our uncertainty is not generating behaviour obviously stronger 
than $(\log_{10} E_{\nu})^2$ for $E_{\nu} \leq 10^{12}$~GeV. 
 
Finally, the fraction of NC events is parametrized by:
\begin{equation}
 \frac{\sigma_{\rm{NC}}}{\sigma_{\rm{NC}}+\sigma_{\rm{CC}}}  = D_1 + D_2 \cdot \ln\left(\varepsilon-D_0\right) 
\end{equation}
with $D_0=1.76$, $D_1=0.252162$ and $D_2=0.0256$.

\subsection{Differential Cross Sections}
\label{sec:dsigmadydx}

\subsubsection{Inelasticity}
\label{sec:dsigmady}

For ease of use in Monte Carlo programs, we describe here a procedure 
for choosing inelasticities that follow the proper energy-dependent 
distributions for energies in the range $10^{7}\leq E_{\nu} \leq 10^{12}$~GeV.
We use the Inverse Transform Method described in~\cite{Amsler:2008zzb}, which requires 
finding a function that describes $d\sigma/dy$ which has an integral
that is invertible.

Due to the steepness of the differential cross section at low values of $y$, 
we divide up the parametrization into two regions in $y$:
\begin{equation}
\frac{d\sigma}{dy} = \left\{ 
\begin{array}{l l}
 Y(C_0,C_1,C_2 ) & \quad \mbox{$0<y<10^{-3}$}\\
Y^{\prime}(C_0^{\prime},C_1^{\prime}) & \quad \mbox{$10^{-3}<y<1$}\\ \end{array} \right. 
\end{equation}
with $Y$ and $Y^{\prime}$ taking the following form:
\begin{equation}
\label{eq:dsigmady_lowy}
Y\left(C_0,C_1,C_2 \right)=
\frac{C_0}{(y-C_1)^{1/C_2}}.
\end{equation}
\begin{equation}
\label{eq:dsigmady_highy}
Y^{\prime}\left(C_0^{\prime},C_1^{\prime} \right)=
\frac{C^{\prime}_0}{y-C_1^{\prime}}.
\end{equation}

For the low $y$ region, the normalized integral of the distribution at $y_0$ is:
\begin{equation}
\label{eq:intdsigmady_lowy} 
I\left(y_0 \right)=
\frac{\int_{y_{\rm{min}}}^{y_0}{Y\left(C_0,C_1,C_2  \right) dy}}
{\int_{y_{\rm{min}}}^{y_{\rm{max}}}{Y \left(C_0,C_1,C_2 \right) dy}}
=\frac{(y_0-C_1)^{(-1/C_2+1)} - (y_{\rm min}-C_1)^{(-1/C_2+1)}}{(y_{\rm max}-C_1)^{(-1/C_2+1)} - (y_{\rm min}-C_1)^{(-1/C_2+1)}}.
\end{equation}
For the high $y$ region, it is:
\begin{equation}
\label{eq:intdsigmady_highy} 
I\left(y_0 \right)=
\frac{\int_{y_{\rm{min}}}^{y_0}{Y^{\prime}\left(C_0^{\prime},C_1^{\prime}  \right) dy}}
{\int_{y_{\rm{min}}}^{y_{\rm{max}}}{Y^{\prime} \left(C_0^{\prime},C_1^{\prime} \right) dy}}
=\left[ \frac{\ln{\left( \frac{y_0-C_1^{\prime}}{y_{\rm{min}}-C_1^{\prime}} \right)}}
{\ln{\left( \frac{y_{\rm{max}}-C_1^{\prime}}{y_{\rm{min}}-C_1^{\prime}} \right)}}\right].
\end{equation}

Notice that Equations~\ref{eq:intdsigmady_lowy} and~\ref{eq:intdsigmady_highy} no longer contain 
the normalization factors $C_0$ and $C_0^{\prime}$.  
For the low-$y$ region, $y_{\rm{min}}=0$ and
$y_{\rm{max}}=10^{-3}$, while 
for the high-$y$ region, $y_{\rm{min}}=10^{-3}$ and
$y_{\rm{max}}=1$.

Since Equations~\ref{eq:intdsigmady_lowy} and~\ref{eq:intdsigmady_highy} 
each represent a cumulative distribution function that 
is invertible, we can use the Inverse Transform Method 
to select values of $y_0$ that follow the
distributions in Equation~\ref{eq:dsigmady_lowy} and~\ref{eq:dsigmady_highy} in each region.  
By setting $I(y_0)$ of Equations~\ref{eq:intdsigmady_lowy} and~\ref{eq:intdsigmady_highy} to a random number 
R between 0 and 1, we can then solve for our choice of $y_0$ 
and obtain for the low $y$ region:
\begin{equation}
\label{eq:choosey0_lowy}
y_0=C_1+\left[R(y_{\rm max}-C_1)^{(-1/C_2+1)}+(1-R)(y_{\rm min}-C_1)^{(-1/C_2+1)} \right]^{C_2/(C_2-1)}
\end{equation} 
and for the high $y$ region:
\begin{equation}
\label{eq:choosey0_highy}
y_0=\frac{(y_{\rm{max}}-C_1^{\prime})^R}{(y_{\rm{min}}-C_1^{\prime})^{R-1}}+C_1^{\prime}
\end{equation}

The parameter $C_1$ itself depends on $\varepsilon$,
and for both regions of $y$, the energy dependent parameter takes
the form: 
\begin{equation}
\label{eq:param_C1}
C_1=A_0- A_1\exp{\left[-\left(\varepsilon-A_2\right)/A_3\right]}.
\end{equation}
In Equation~\ref{eq:param_C1}, all parameters 
are primed when describing the high-$y$ region. 
The numerical values of the parameters in Equation~\ref{eq:param_C1},
summarized in Table~\ref{tab:param}, were obtained from fits of the 
parametrizations to the theoretical calculations. 

In the low region, the parameter $C_2$ is also energy dependent:
\begin{equation}
\label{eq:param_C2}
C_2=B_0+ B_1 \cdot \varepsilon
\end{equation}
where $B_0=2.55$ and $B_1=-0.0949$ for all interaction types.

The fraction of the cross section occupying the low-$y$ region is given by:
\begin{equation}
\label{eq:frac}
f(\varepsilon)=F_0\cdot\sin{[F_1 \cdot (\varepsilon-F_2)]}
\end{equation}
with $F_0=0.128$, $F_1=-0.197$ and $F_2=21.8$ for all interaction types.

\begin{figure}
    \includegraphics[width=3.4in]{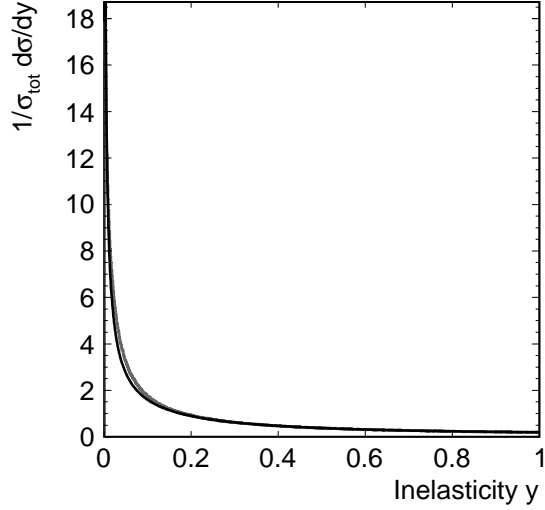}
\caption{\label{fig:dy} Comparison of the $\nu N$ CC inelasticity distributions
from our theoretical calculation (black line) and 
the distribution obtained from the parametrization procedure described
in Section~\ref{sec:dsigmady} (gray histogram).  }
\end{figure}

\begin{figure}
    \includegraphics[width=3.4in]{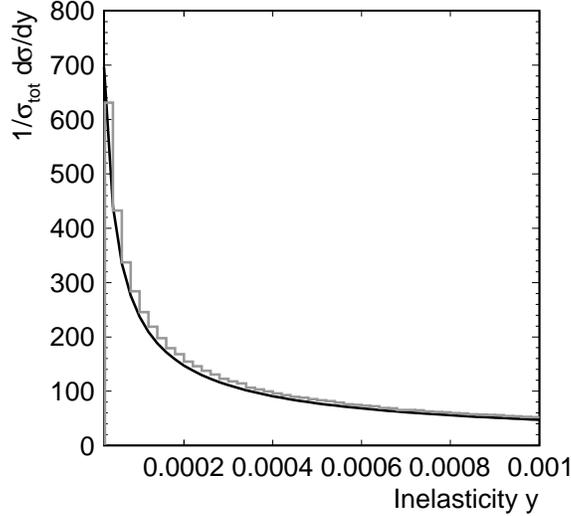}
\caption{\label{fig:dy_low} Same as Fig.~\ref{fig:dy}, but for low-$y$ region.}
\end{figure}

To summarize, for an interaction of a given type ($\nu N$ or $\bar{\nu} N$, CC or NC) 
at an energy $\varepsilon$, one can find an inelasticity $y_0$ chosen from the 
appropriate distribution through the following steps:
\begin{itemize}
\item{Choose a random number $R_1$ between 0 and 1 and if $R_1<f(\epsilon)$ 
(see Equation~\ref{eq:frac}), then the event lies in the low-$y$ region.  Otherwise, 
it is in the high-$y$ region.}
\item{Obtain the value of $C_1^{\prime}$, or $C_1$ and $C_2$, depending on the $y$ region
and the event type, using Equations~\ref{eq:param_C1}, Equation~\ref{eq:param_C2}
and Table~\ref{tab:param}.}
\item{Choose a new random number $R_2$ and insert $R=R_2$ along with the
parameters obtained in the previous step
 into Equation~\ref{eq:choosey0_lowy} 
or Equation~\ref{eq:choosey0_highy}
to obtain $y_0$.}
\end{itemize}
Figs.~\ref{fig:dy} and~\ref{fig:dy_low} show 
the calculated $y$ distributions in each region compared with
the event distributions generated from this procedure for 
$\nu N$, CC events at $\varepsilon=12$.  


For all event types and energies
in the range $7\leq \varepsilon \leq 12$, this procedure will give
$y$ distributions whose mean value differs from the theoretical
calculation by at most 3.7\% in the energy range $E_{\nu}=10^{7}-10^{12}$~GeV.  
Recall that the inelasticity is the fraction of neutrino 
energy carried away by the hadronic shower
and therefore these uncertainties on the mean hadronic energy 
imply less than 1\% uncertainties on the mean energy of the final state
lepton. 
The RMS of the distributions are 0.05-0.07 in the
same energy range and the difference in RMS values between the 
model and the calculation is no more than 2.6\% for all interaction
types except the $\bar{\nu}N$ CC events, whose uncertainties on
the RMS values do not exceed 8.3\%.  In~\cite{Bulmahn:2009ub}, the authors give
a parametrization of the inelasticity distributions in the
neutrino energy range 50~GeV$<E_{\nu}<10^{12}$~GeV which are
within 15\% agreement with calculations using the CTEQ6
parton distribution functions.

Sample code for generating energy dependent inelasticity distributions for all interaction
types according to the prescription laid out in this paper can be found at:\\
{\tt http://www.physics.ohio-state.edu/$\mathtt{\sim}$connolly/crosssections/y.html}.

\begin{table}
    \caption{Coefficients that go into calculating $C_1$ and $C_1^{\prime}$ 
in Equation~\ref{eq:param_C1} for parametrizing the inelasticity distributions. \label{tab:param}}
   \begin{center}
      \begin{tabular}{p{1.8cm}p{1.5cm}p{1.5cm}p{1.5cm}p{1.5cm}} \hline \hline
\multicolumn{5}{c}{low $y$} \\
 & $A_{0}$ & $A_{1}$ & $A_{2}$ & $A_{3}$    \\ \hline 
  & 0.0 & 0.0941 & 4.72 & 0.456   \\ \hline
\multicolumn{5}{c}{high $y$}\\ 
& $A^{\prime}_{0}$ & $A^{\prime}_{1}$ & $A^{\prime}_{2}$ & $A^{\prime}_{3}$    \\ \hline 
$\bar{\nu}N$ CC & -0.0026 & 0.085 & 4.1 &\multirow{2}{*}{1.7}  \\  
$\nu N$ CC & -0.008 & 0.26 & 3.0 &  \\ \hline
$\bar{\nu}N$ NC & \multirow{2}{*}{-0.005} & \multirow{2}{*}{0.23} & \multirow{2}{*}{3.0} & \multirow{2}{*}{1.7} \\  
$\nu N$ NC  & & & & \\ \hline \hline
\end{tabular}
    \end{center}
\end{table}

\subsubsection{Bjorken-$x$}

\begin{figure}
    \includegraphics[width=3.4in]{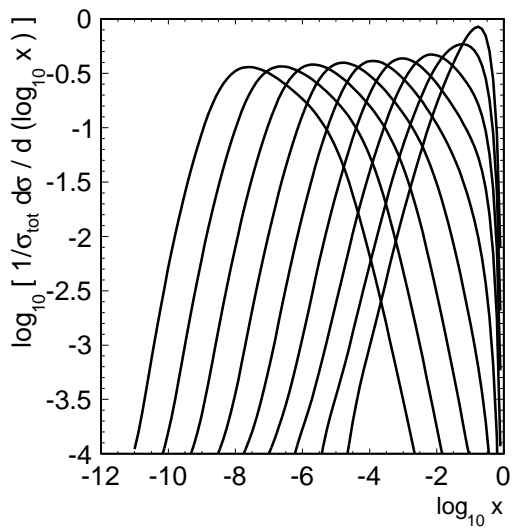}
\caption{\label{fig:dx} Normalized $x$ distribution for 
neutrino interactions for each decade in energy from
$E_{\nu}=10^4$ to $10^{12}$~GeV.  As neutrino energy increases, 
the mean value of $\log_{10} x$ decreases.}
\end{figure}

The fraction of momentum carried by a parton within a nucleon is 
called the Bjorken-$x$.  UHE neutrino cross sections 
include contributions from PDFs in the region of $x$ 
that is lower than the region above $10^{-4}-10^{-5}$ 
accessible in the perturbative regime by HERA experiments~\cite{:2009wt}.  
Fig.~\ref{fig:dx} shows the distributions of 
$\left(1/\sigma_{\rm{tot}}\right) d\sigma/d(\log_{10}x)$ for energies between
$10^{4}$ and $10^{12}$~GeV. As one can see the cross section starts to become very 
sensitive to the $x$-range below the extent of the HERA data constraint at $E_{\nu}
\sim 10^9$ GeV.  

\subsection{Correlations}
\label{sec:correlations}
Since we quote uncertainties on the $\nu N$ cross sections continuously across
the energy range, we also include the correlations between the uncertainties at
different energies using the prescription laid out in~\cite{Nadolsky:2008zw}.
For completeness we briefly summarize the procedure here.

Each cross section value calculated in this paper is the sum of contributions
from $N$ different orthogonal eigenvectors which are the result of diagonalizing
the parameters of the PDFs. Considering two different cross sections $X$ and $Y$,
the correlation between their uncertainties is denoted $\cos{\varphi}$ and is 
given by:
\vspace{0.2in}
\begin{equation}
\begin{array}{rcl}
\cos{\varphi} &= &\frac{\vec{\Delta} X \cdot \vec{\Delta} Y}{\Delta X \Delta Y}\\
&& = \frac{1}{4\Delta X\Delta Y}\sum_{i=1}^{N} \left( X_i^{(+)}-X_i^{(-)}\right) \left( Y_i^{(+)}-Y_i^{(-)} \right)
\end{array}
\end{equation}
where
 \begin{equation}
\Delta X = \left| \vec{\Delta} X \right| = \frac{1}{2} \sqrt{\sum_{i=1}^{N} \left( X_i^{(+)}-X_i^{(-)} \right)^2}.  
\end{equation}

Here, $X_i^{(+)}$ and $X_i^{(-)}$ are the upper and lower bounds on the contribution to
$X$ from the $i^{\rm{th}}$ eigenvector.  For a maximum correlation, $\cos{\varphi}=1$,
for an anticorrelation, $\cos{\varphi}=-1$ and for two quantities that are uncorrelated,
$\cos{\varphi}=0$.

\begin{figure}
    \includegraphics[width=3.4in]{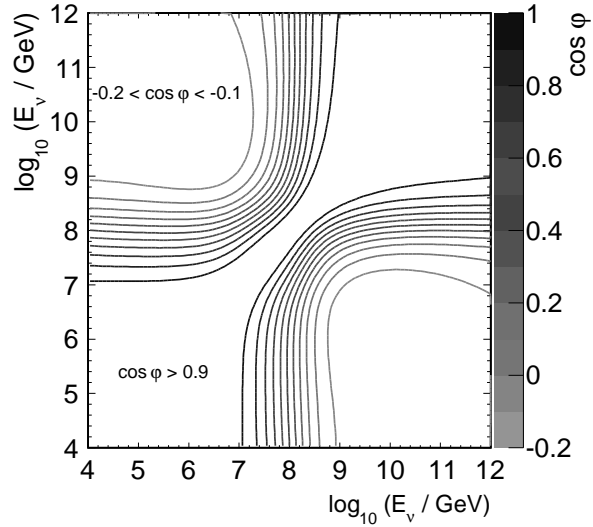}
\caption{\label{fig:corr2_cc} Correlation factor $\cos{\varphi}$ between the
charged current $\nu N$ cross sections at the energy on
the abscissa and the energy on the ordinate.  }
\end{figure}
In Fig.~\ref{fig:corr2_cc}, we plot $\cos{\varphi}$ for all 
energy pairs in the
range $10^4-10^{12}$~GeV for 
charged current $\nu N$ cross sections.  The analogous plots 
for $\bar{\nu}N$ and
neutral current cross sections look similar.
There are very strong correlations among cross sections at energies above
$10^9$~GeV, and then again below $10^7$~GeV, with little correlation
between energies in different regions.  There are three reasons for this, all 
associated with the fact that in the lower energy region the cross section has 
a high proportion of its contribution from $x \sim 0.01$ or above, while in the 
higher energy region most of the contribution is from $x$ values lower than this. 
First, from sum rules in the PDFs, there is a crossing point where changes in PDFs
become anticorrelated, i.e. for any change the PDFs, they tend to increase
below this $x$ and decrease above this $x$ or {\it vice versa}. 
For high energy scales this is at $x\approx0.01$. 
Second, this also happens to be the $x$ where there is a very large
amount of accurate HERA data also tending to fix the PDFs. 
Thirdly, $x=0.01$ is a transition point at which for higher
$x$ the dominant contributions are from valence quarks but for 
lower $x$ the sea quarks dominate, which are gluon driven.

\section{Cross Section Constraints}
\label{part:meas}

\subsection{Motivation}
\label{sec:motivation}

The upper (lower) bounds on the Standard Model $\nu N$ cross sections 
differ from the central values by more than approximately 20\% (15\%)
for neutrino energies above $10^9$~GeV.  Above $10^{10}$~GeV,
the uncertainties are more than approximately 50\% (30\%).
We would like to constrain the cross sections at the highest energies 
because in that region neutrino experiments could be sensitive to new 
physics scenarios. Fortunately, there is a near guaranteed flux of   
neutrinos in the UHE energy regime from GZK interactions, but this so-called ``cosmogenic'' 
neutrino flux has large theoretical uncertainties associated with it.

Here we outline a technique~\cite{Connolly:2006gh} 
to constrain the UHE $\nu N$ cross sections
that is {\it independent of the incident neutrino flux} 
through the measured zenith angle distributions with a
subterranean detector such as IceCube, ARA or ARIANNA~\cite{Achterberg:2006md,HoffmanArena,Gerhardt:2010js}.  
The latter two experiments are currently in the first stages of deploying prototype
detectors and will be focused on the UHE regime.  Here, we focus
on the ARA detector as an example, but our results are general to
any subterranean neutrino detector with similar capabilities
in energy measurement and reconstruction.

The sensitivity of neutrino detectors to $\nu N$ cross sections due to
earth absorption has been addressed elsewhere in varying degrees~\cite{Kusenko:2001gj,Hooper:2002yq,Anchordoqui:2010hq,Anchordoqui:2005ey,Anchordoqui:2005pn,Borriello:2007cs,Romero:2010zza,Hussain:2006wg,Hussain:2007ba}.
This is the only study that uses the full zenith angle distribution 
in the UHE regime to make hard predictions for the expected constraints on
models with extra dimensions.

We note a few assumptions made for this study. First, we
assume the incident flux is entirely neutrinos with no anti-neutrino
component.  Above $10^8$~GeV, the cross sections for neutrinos
and anti-neutrinos differ by no more than about 6\%.
Second, we assume that all of the interactions occur 
at the same depth, $d=250$~m, and that it is precisely known.
In an actual data analysis one would modify the $dn/d\cos{\theta_z}$ 
distributions so that $d$ is the measured depth, just as we use the
measured energy for each event.  The three-dimensional vertex resolution of the
ARA detector is expected to be of order 10's of meters. One can show that 
a depth uncertainty given by $\delta d$ changes $dP/d\cos{\theta_z}$
by a fraction of order $\delta d/L$.  This only approaches of order 10\%
at the highest cross sections probed in this paper, approximately $10^{-28.5}$~cm$^2$.
We also assume that the detector efficiency is flat in zenith angle
$\theta_z$ and that the neutrino flux is isotropic.  An analysis would
fold in the measured detector efficiency, and neutrinos from any
 discovered sources could be removed from the sample.
Finally, we assume that the energy resolution
$\Delta \log_{10}E_{\nu}=0.4$, which
is dominated by the uncertainty in 
inelasticity, is appropriate for both neutrino 
and black hole interactions.

\subsection{ARA}
\label{sec:ARA}
The Askaryan Radio Array (ARA) is a nascent neutrino detector
near the South Pole.  It is an array of radio antennas 
deployed deep in the ice, designed to measure the radio
Cerenkov pulse from UHE neutrinos~\cite{Askaryan:1962,Askaryan:1965,AlvarezMuniz:1997sh,Saltzberg:2000bk}.
If expanded to become a precision measurement,
observatory class array of 300 to 1000 km$^2$ area,
ARA will be capable of measuring hundreds of cosmogenic neutrinos
per year. The
first ARA testbed station was deployed
in the 2010–2011 austral summer and the first ARA stations will be deployed in 2011-2012.

{

For the purpose of generating mock ARA data for this study, 
we use energy-dependent relative effective
areas (the overall scale is not used) 
derived by inverting the projected flux limits for
a 37-station array in~\cite{HoffmanArena}.  
The stations are arranged on a triangular grid
with the array forming a hexagon.  Each station is made of
three ``strings'' deployed vertically in the ice,
each holding two pairs of vertically and horizontally
polarized antennas that sit at 200~m depth, for
a total of 12 antennas per station.  The trigger requires
5 out of 12 stations measure a pulse that exceeds 
3.5 times the expected noise level.
We assume an energy resolution of $\Delta\log_{10}{E_{\nu}}=0.4$.
The resolution on the neutrino $\theta_z$
is expected to be approximately $2^{\circ}$ for the 
events (approximately 80\%)
detected by one station only~\cite{gorham}.

Our projected constraints will not depend on the exact size
of the ARA detector.   This is because
the expected limits will be quoted for a specific number
of neutrino events measured in the detector, whether that
came from a weak flux measured with a large detector
or a strong flux measured with a smaller detector.
Nor do the projected constraints depend 
strongly on the energy threshold of the
experiment, as long as there is 
at least a crude energy measurement capability.
We will quantify these statements at the conclusion
of Section~\ref{sec:limits}.
In addition, our conclusions are not specific to the ARA
experiment, and would be qualitatively similar for
any subterranean detector with sensitivity to UHE neutrinos.

\subsection{A Flux-Independent Technique for Measuring the UHE Neutrino-Nucleon Cross Section }
\label{sec:technique}

Consider a neutrino (see Figure~\ref{fig:drawing8}) 
that interacts in a subterranean detector such as ARA  at depth $d$ and 
zenith angle $\theta_z$, the angle from vertical
of the direction of origin of each incident neutrino.  The neutrino travels a distance
$D$ through the earth of radius $R$ before reaching its interaction point.  From
Figure~\ref{fig:drawing8} and using the law of sines,
\begin{equation}
\frac{\sin{\phi_1}}{R-d}=\frac{\sin(\pi-\theta_z)}{R}.
\end{equation}
which gives
\begin{equation}
\sin{\phi_1}=\frac{R-d}{R}\sin{\theta_z}.
\end{equation}
Then since $\phi_2=\pi/2-\phi_1$,
\begin{equation}
\sin{\phi_2}=\cos{\phi_1}=\sqrt{1-\sin^2{\phi_1}}=\sqrt{1-\left(\frac{R-d}{R} \right)^2 \sin^2{\theta_z}}
\label{eq:sinphi2}
\end{equation}
Then the distance traveled through the earth by the neutrino is:
\begin{equation}
D = (D-x) + x = R\sin{\phi_2}+(R-d) \sin\left( {\theta_z-\frac{\pi}{2}}\right)
\label{eq:D1}
\end{equation}
Inserting Eq.~\ref{eq:sinphi2} into Eq.~\ref{eq:D1} we find:
\begin{equation}
D = R\sqrt{1-\left(\frac{R-d}{R} \right)^2 \sin^2{\theta_z}} +(R-d) \sin\left( {\theta_z-\frac{\pi}{2}}\right)
\label{eq:D2}
\end{equation}
Then, taking $d<<R$ and replacing $\sin{\theta_z}$ with $\sqrt{1-\cos^2{\theta_z}}$, we find
\begin{equation}
D=\sqrt{(R^{2}-2Rd)\cos^2{\theta_z}+2Rd} - (R-d)\cos{\theta_z}.
\end{equation}
If the detection efficiency is uniform in $\theta_z$, then for
a neutrino with energy $E_{\nu}$, the probability
 distribution in $\theta_z$ for a detected interaction is given by~\cite{GonzalezGarcia:2007gg}:
\begin{equation}
\label{eq:dPdcostheta}
\frac{dP}{d\cos{\theta_z}}\left(E_{\nu}\right)=A\cdot \exp{\left(-\frac{D}{L\left(E_{\nu} \right)}\right)}=
A\cdot \exp\left( 
-\frac{\sqrt{(R^{2}-2Rd)\cos^2{\theta_z}+2Rd} - (R-d)\cos{\theta_z} }{L\left(E_{\nu} \right) }    
\right) 
\end{equation}
where $R$ is the radius of the earth, $A$ is a constant 
that sets the total probability to unity, 
and $L\left(E_{\nu} \right) $ is the interaction length 
for a neutrino of energy $E_{\nu}$ along its path through the earth, given by:
\begin{equation}
L\left(E_{\nu} \right)=\frac{M_N}{ \sigma(E_{\nu})\left< \rho \right>_{\theta_z} } 
\end{equation}
where $M_N$ is the nucleon mass, $\sigma(E_{\nu})$ is the $\nu N$ cross section
at energy $E_{\nu}$ and $\left< \rho \right>_{\theta_z} $ is the mean density
averaged over the distance travelled by the neutrino at the given $\theta_z$.
The expected $\theta_z$ distribution from a sample of $n$ measured neutrino
interactions with measured energies $\tilde{E}_{\nu}^i$ 
will be the sum of $n$ different $(dP/d\cos{\theta_z})_i$, 
so that the resulting expected 
$\theta_z$ distribution is then given by:
\begin{equation}
\label{eq:dndcostheta}
\frac{dn}{d\cos{\theta_z}}_{\rm{exp}}= \sum_{i=1}^{n} A_i\cdot \exp\left( 
-\frac{\sqrt{(R^{2}-2Rd)\cos^2{\theta_z}+2Rd} - (R-d)\cos{\theta_z} }{L\left(\tilde{E}_{\nu}^i \right) }    
\right).
\end{equation}
Note that due to the difference between measured neutrino energies $\tilde{E}_{\nu}^i$
and true energies $E_{\nu}^i$, the expected distribution will differ from the true one.
In Fig.~\ref{fig:thetanu}, we plot $dP/d\cos{\theta_z}$ for
a few monoenergetic distributions.  Notice the breaks in the distribution
due to the neutrino trajectory intersecting the Earth's core and mantle,
derived using a simple three-layer model of the Earth's interior.
Due to this structure at lower energies, this technique could lead to a measurement
of the Earth's density profile with a km$^3$ neutrino experiment such as IceCube 
that is independent of the traditional
techniques used by geologists~\cite{GonzalezGarcia:2007gg}.

\begin{figure}
   \includegraphics[width=3.4in]{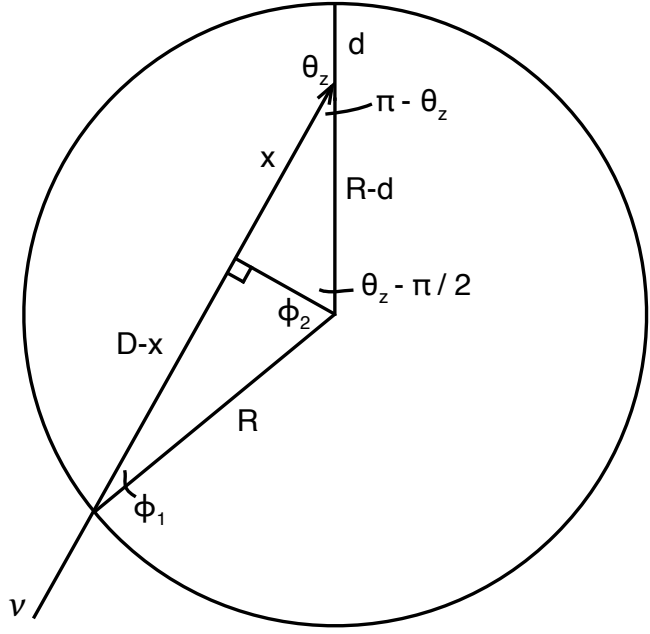}
    \caption{ Diagram showing neutrino incident on the earth used in derivation
of Eq.~\ref{eq:dPdcostheta}.
\label{fig:drawing8}}
\end{figure}

\begin{figure}
   \includegraphics[width=3.4in]{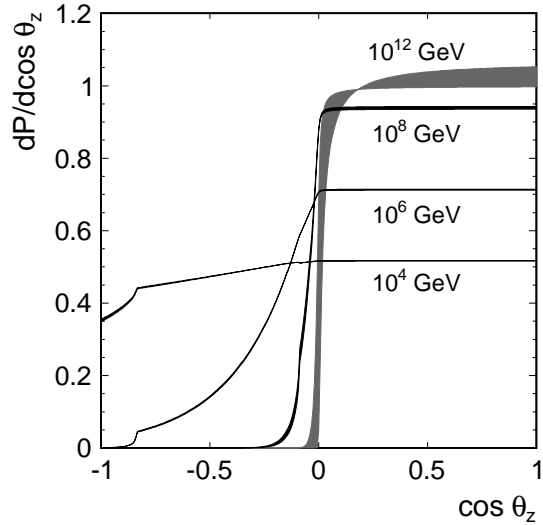}
    \caption{ Probability distributions in zenith angle for 
select neutrino energies.  
    The width of each band is due to the cross section uncertainties reported in this paper.  
    The kinks in the distributions at $\cos{\theta_z}=-0.1$ and $\cos{\theta_z}=-0.8$ 
    are due to the neutrino paths reaching the earth's mantle and core, respectively.
\label{fig:thetanu}}
\end{figure}

\subsection{Models with Extra Space-Time Dimensions}
\label{sec:edm}

There are a class of models for physics beyond the Standard Model
that contain extra space-time dimensions~\cite{Antoniadis:1998ig,ArkaniHamed:1998rs,Anchordoqui:2003jr,AlvarezMuniz:2002ga}.  
These models are motivated
by the need to resolve what is known as 
the Hierarchy Problem in particle physics, where the dramatically
different energy scales for electroweak symmetry breaking and quantum 
gravity lead to a need for fine tuning of terms in the calculation 
of the Higgs mass.

In these extra-dimensional models (EDMs), 
the energy scale at which gravity dominates, $M_D$,
is reduced to of order 1~TeV, just above the electroweak scale.
The weakness of gravity in our 3+1 dimensional world is a 
consequence of its propagation in additional dimensions.  The number
of dimensions in the model beyond the four known space-time dimensions  
is denoted $N_D$.

Interactions at energies at or above the reduced Planck mass
 lead to the production of micro-black holes, and this additional channel
causes cross sections to be enhanced.  The minimum black hole
mass is given by $M_{\rm{BH}}^{\rm{min}}~=~x_{\rm{min}}M_D$, where $x_{\rm{min}}$ is
a parameter in the model.  
Fig.~\ref{fig:bh} shows
the predicted $\nu N$ cross sections for a few EDMs, from~\cite{AlvarezMuniz:2002ga}, 
compared to the SM cross sections
calculated in this paper.  

Tevatron experiments CDF and D0
have already set lower
limits on $M_{\rm{D}}$ in the range of approximately 1-1.6~TeV
with between 2 and 7 extra dimensions~\cite{Aaltonen:2008hh,:2009mh}.
A recent paper by the CMS collaboration 
places constraints on black hole production at the LHC based on
35~pb$^{-1}$ of data at center-of-mass energy of 
7~TeV~\cite{Khachatryan:2010wx}.  For 
$x_{\rm{min}}=1$, they exclude models with $n$ up to 6 for 
$1.5<M_{\rm{D}}<3.5$~TeV. They were not sensitive to models with
$x_{\rm{min}}=3$ in the range of $M_{\rm{D}}$ considered.
In addition, neutrino experiments have constrained UHE cross sections, but
in a way that depends on a model for the neutrino flux~\cite{Yoshida:2010kp}.
With the technique described here, neutrino experiments would 
be setting limits on EDMs that are competitive with collider
experiments and independent of a flux model.

\begin{figure}
    \includegraphics[width=3.4in]{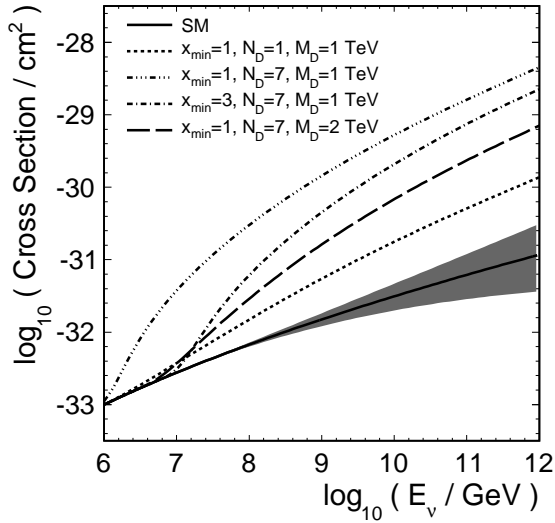}
\caption{\label{fig:bh} Cross sections for $\nu N$ interactions in 
models with extra space-time
dimensions compared with the SM $\nu N$ cross sections.  The gray band surrounding the
SM cross sections are the uncertainties presented in this paper.   }
\end{figure}

In Fig.~\ref{fig:thetanu_bh}, we plot the expected $\theta_z$ 
distribution for 100~events
measured in ARA using
cross sections from the SM compared to ones from a few select EDMs
and a bin width of $\Delta \cos{\theta_z}=0.1$.
The shape of the true energy spectrum of the 100~events is the product of an incident
flux spectrum and an energy dependent effective area for the detector.
Here we assume a neutrino flux from the GZK process 
as in~\cite{Engel:2001hd}, and the ARA effective areas
derived from~\cite{HoffmanArena} as described in Section~\ref{sec:ARA}.
\begin{figure}
   \includegraphics[width=3.4in]{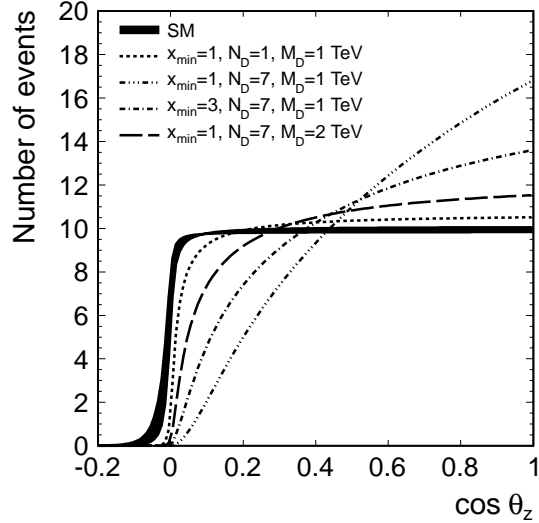}
    \caption{Predicted $\cos{\theta_z}$ distributions for 100 neutrino events 
observed by ARA for the SM and the four different EDMs shown in Figure~\ref{fig:bh}.\label{fig:thetanu_bh}.  
The bin width is taken to be $\Delta \cos{\theta_z}=0.1$.}
\end{figure}

\subsection{Projected Constraints}
\label{sec:limits}

In order to assess the sensitivity of a future ARA detector 
to EDMs, we generate many pseudoexperiments, 
and compare the resulting
pseudodata distributions 
to the predicted ones
for the signal
and null hypothesis respectively. 
For an expected number of events $N_{\rm{exp}}$, the number of events
observed in a given pseudoexperiment is given by $n_{\rm{p}}$
and is Poisson distributed with
mean $N_{\rm{exp}}$.
The data in the $i^{\rm{th}}$ 
bin in $\cos{\theta_z}$ is denoted $n_{\rm{p},i}$.
The mean number of events predicted in the $i^{th}$ bin 
centered on $\cos{\theta_z}=c_0$ is
\begin{equation}
\mu_i=\frac{dn_{\rm{p}}}{d\cos{\theta_z}} \left( c_0  \right) \cdot \Delta \cos{\theta_z}
\end{equation}
where $dn_{\rm{p}}/d\cos{\theta_z}$ is constructed for each pseudoexperiment
using Equation~\ref{eq:dndcostheta}.
Note that through the $L(\tilde{E}_{\nu}^i)$'s in Equation~\ref{eq:dndcostheta},
$dn_{\rm{p}}/d\cos{\theta_z}$ depends on the model hypothesis,
whether that be the SM or an EDM.  Also note that we only use the
measured energies, selected from the incident flux spectrum and
then smeared according to detector resolution of $\Delta \log{E_{\nu}}=0.4$.
 The bin width is
$\Delta\cos{\theta_z}=0.1$, or $\Delta\theta_z\approx 5.7^{\circ}$, 
which is greater than the expected resolution
of the ARA array in $\theta_z$.

We use the following ratio of Poisson probabilities to discriminate between
the two hypotheses~\cite{CasalLarana:2010zz}:  
\begin{equation}
Q=\frac{P_{\rm{poiss}}(\rm{data}~|~\rm{EDM~truth})}{P_{\rm{poiss}}(\rm{data}~|~\rm{SM~truth})}
\end{equation}
where
\begin{equation}
P_{\rm{poiss}}(\rm{data}~|~\rm{EDM~truth})=\prod_{i=1}^{N} \frac{\mu_{\rm{EDM},i}^{n_i}e^{- \mu_{\rm{EDM},i}}}{n_i!}
\end{equation}
and
\begin{equation}
P_{\rm{poiss}}(\rm{data}~|~\rm{SM~truth})=\prod_{i=1}^{N} \frac{\mu_{\rm{SM},i}^{n_i}e^{- \mu_{\rm{SM},i}}}{n_i!}.
\end{equation}
Here, $N$ is the number of bins, and $n_i$ is the number of events measured in the $i^{th}$ bin.  The number of events expected in a bin from an extra-dimensional model is $\mu_{\rm{EDM}}$ and the number expected in the Standard Model is $\mu_{\rm{SM}}$.
Then, we find
\begin{equation}
\label{eq:like}
-2\ln{Q}=-2\left[ \sum_{i=1}^{N}n_i \ln\left({\frac{\mu_{\rm{EDM},i}}{\mu_{\rm{SM},i}}}\right)-\mu_{\rm{EDM},i}+\mu_{\rm{SM},i} \right]
\end{equation}

The parameters for the EDM models are defined in Section~\ref{sec:edm}.
Equation~\ref{eq:like} is evaluated separately 
for pseudodata $\vec{n_{\rm{p}}}$ 
generated assuming SM and EDM truths, giving 
a different $-2\ln{Q}$ distribution for each.
According to the Neyman-Pearson lemma, this likelihood ratio is 
the test statistic with the most discriminating power~\cite{NeymanPearson}.

We estimate the constraints on EDMs expected to be set by the
full scale
ARA detector described in Section~\ref{sec:ARA} 
depending on the observed number of events $n_{\rm{p}}$.
For the expected limit, 
we first consider the median value $-2\ln{Q}_{50}$ of 
the $-2\ln{Q}$ distribution from
SM pseudoexperiments.
The subscript denotes the percentage of 
SM pseudoexperiments with lower values of $-2\ln{Q}$.
Then, for a given EDM model,
the $p$ value is the fraction of
EDM pseudoexperiments with $-2\ln{Q}<-2\ln{Q}_{50}$.
Then, on average we can expect that 
a model can be excluded with percentage 
confidence level $CL=100\times(1-p)$.

The observed limit will differ from that expected due to fluctuations
in the data, as reflected by the width of the $-2\ln{Q}$ distribution
for SM pseudoexperiments. Therefore, we quote a range for the expected $CL$s
by calculating $p$ values for $-2\ln{Q}_{16}$ and $-2\ln{Q}_{84}$, so 
that 68\% of pseudoexperiments would give confidence levels in the range.
Fig.~\ref{fig:loglik} shows the distributions in $-2\ln{Q}$ for
the SM and one EDM hypothesis, for
100 events observed in ARA.  We mark the points along the
abscissa $-2\ln{Q}_{16}$, $-2\ln{Q}_{50}$ and $-2\ln{Q}_{84}$.
\begin{figure}
    \includegraphics[width=3.4in]{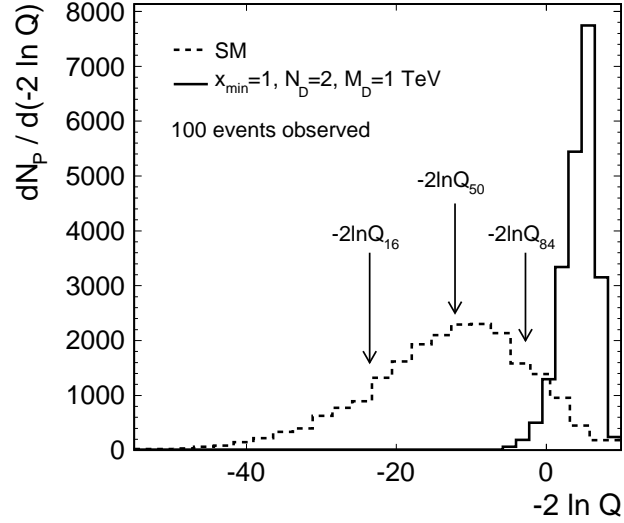}
\caption{\label{fig:loglik}Comparison of likelihood ratio $-2\ln{Q}$ from 
$N_{\rm{P}}$ pseudoexperiments generated assuming SM and EDM truth.  
The region between $-2\ln{Q}_{16}$ and $-2\ln{Q}_{84}$ contains 
68\% of the SM pseudoexperiments,
 with the median at $-2\ln{Q}_{50}$. }
\end{figure}

Fig.~\ref{fig:limits} shows the expected CLs for constraining EDMs as a 
function of the number of events observed with ARA.  The black bands show the
variation in the expected constraints brought about by the uncertainties
on the SM cross sections presented here.
The gray bands show the range of expected limits due to variations in
the data; sixty-eight percent of the pseudoexperiments
give constraints that lie in the gray region.
The gray bands are centered on the curve corresponding to the central
value standard model cross sections. 
 Conservatively taking the SM upper bounds to be the true
cross sections, for 100 events observed with ARA, the mean expectation
is to exclude the following models:  $x_{\rm{min}}~=~1,~M_{\rm{D}}=1,~N_{\rm{D}}\geq 2$;
$x_{\rm{min}}~=~3,~M_{\rm{D}}=1,~N_{\rm{D}}\geq3$;
$x_{\rm{min}}~=~1,~M_{\rm{D}}~=~2,~N_{\rm{D}}\geq3$.  
For $x_{\rm{min}}~=~3,~M_{\rm{D}}~=~2$, 110 events would be needed to be predicted
to exclude $N_{\rm{D}}=7$.

\begin{figure}
\includegraphics[width=6.0in]{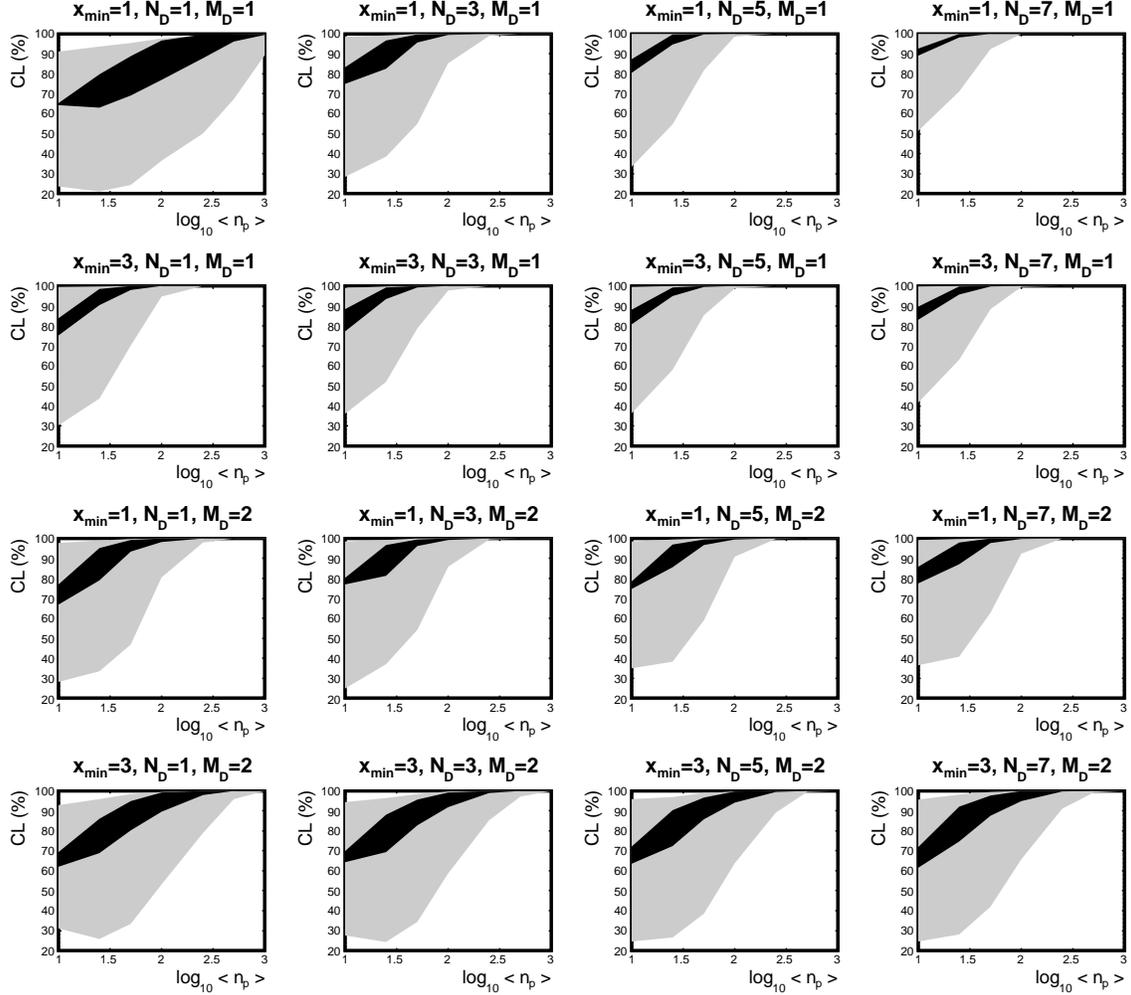}
\caption{\label{fig:limits} Predicted confidence levels for excluding 
selected EDMs
with an ARA experiment, depending on the number of events observed.
The upper (lower) 
edges of the black bands in the figure
show the range of expected limits taking the SM cross section
to be their lower (upper) bounds as presented in this paper.
The gray bands show the uncertainty due to 
pseudoexperiment statistics, with 68\% of experiments
expected to result in limits that lie within the gray region 
(using the central values for the SM cross section as the true cross sections).}
\end{figure}


We have checked that these projected limits are robust
to changes in the neutrino spectrum as well as
details of the sensitivity of the experiment.
We have calculated the expected constraints with an $E^{-3}$
neutrino spectrum, and also after shifting the ARA 
effective areas up and down in energy by a factor of 3.
All of these changes have less of an effect on the expected
constraints than the cross section uncertainties (the effect
of which are depicted in the black bands of Fig.~\ref{fig:limits}).

\section{Conclusions}
We have presented new calculations of the $\nu N$ and $\bar{\nu} N$,
CC and NC cross sections in the neutrino energy range
$10^4<E_{\nu}<10^{12}$~GeV using the MSTW~2008 PDFs, along with
their PDF uncertainties.  The cross section values are consistent
with those reported in previous publications but the uncertainties 
presented here are significantly larger. This difference is due to 
a two-parameter model for the gluon parton distribution at very small 
$x$ used by MSTW~2008 arising from a fit to the global data, which leads 
to a wide range of allowed values for the gluon contribution at low-$x$.
For ease of use in Monte Carlo simulations, we have provided
parametrizations of the cross sections for each event type, the
fraction of each type and the cross section uncertainty bounds in the energy 
range. In addition, we have outlined a procedure for generating $y$ values
for a neutrino sample with $E_{\nu}\geq 10^7$~GeV 
using the Inverse Transform Method.

Finally, we present a technique for constraining UHE cross sections
with a next-generation subterranean neutrino experiment.  Using ARA
as an example, we have shown that with 100 events observed,
neutrino experiments set constraints
on extra-dimensional models that are competitive with 
those set by collider experiments.

\begin{acknowledgments}
We are grateful to Amanda Cooper-Sarkar, Subir Sarkar and Grame Watt
for helpful
discussions, and Doug McKay
for his valuable suggestions which improved the 
paper.
 We would also like to thank the Royal Society and
the Science and Technology Facilities Council (STFC) for supporting 
this research.
\end{acknowledgments}


\begin{thebibliography}{50}%
\makeatletter
\providecommand \@ifxundefined [1]{%
 \@ifx{#1\undefined}
}%
\providecommand \@ifnum [1]{%
 \ifnum #1\expandafter \@firstoftwo
 \else \expandafter \@secondoftwo
 \fi
}%
\providecommand \@ifx [1]{%
 \ifx #1\expandafter \@firstoftwo
 \else \expandafter \@secondoftwo
 \fi
}%
\providecommand \natexlab [1]{#1}%
\providecommand \enquote  [1]{``#1''}%
\providecommand \bibnamefont  [1]{#1}%
\providecommand \bibfnamefont [1]{#1}%
\providecommand \citenamefont [1]{#1}%
\providecommand \href@noop [0]{\@secondoftwo}%
\providecommand \href [0]{\begingroup \@sanitize@url \@href}%
\providecommand \@href[1]{\@@startlink{#1}\@@href}%
\providecommand \@@href[1]{\endgroup#1\@@endlink}%
\providecommand \@sanitize@url [0]{\catcode `\\12\catcode `\$12\catcode
  `\&12\catcode `\#12\catcode `\^12\catcode `\_12\catcode `\%12\relax}%
\providecommand \@@startlink[1]{}%
\providecommand \@@endlink[0]{}%
\providecommand \url  [0]{\begingroup\@sanitize@url \@url }%
\providecommand \@url [1]{\endgroup\@href {#1}{\urlprefix }}%
\providecommand \urlprefix  [0]{URL }%
\providecommand \Eprint [0]{\href }%
\providecommand \doibase [0]{http://dx.doi.org/}%
\providecommand \selectlanguage [0]{\@gobble}%
\providecommand \bibinfo  [0]{\@secondoftwo}%
\providecommand \bibfield  [0]{\@secondoftwo}%
\providecommand \translation [1]{[#1]}%
\providecommand \BibitemOpen [0]{}%
\providecommand \bibitemStop [0]{}%
\providecommand \bibitemNoStop [0]{.\EOS\space}%
\providecommand \EOS [0]{\spacefactor3000\relax}%
\providecommand \BibitemShut  [1]{\csname bibitem#1\endcsname}%
\let\auto@bib@innerbib\@empty
\bibitem [{\citenamefont {Berezinsky}\ and\ \citenamefont
  {Zatsepin}(1969)}]{Berezinsky:1969zz}%
  \BibitemOpen
  \bibfield  {author} {\bibinfo {author} {\bibfnamefont {V.~S.}\ \bibnamefont
  {Berezinsky}}\ and\ \bibinfo {author} {\bibfnamefont {G.~T.}\ \bibnamefont
  {Zatsepin}},\ }\href@noop {} {\bibfield  {journal} {\bibinfo  {journal}
  {Phys. Lett.}\ }\textbf {\bibinfo {volume} {B28}},\ \bibinfo {pages} {423}
  (\bibinfo {year} {1969})}\BibitemShut {NoStop}%
\bibitem [{\citenamefont {Berezinsky}\ and\ \citenamefont
  {Zatsepin}(1970)}]{Berezinsky:1970}%
  \BibitemOpen
  \bibfield  {author} {\bibinfo {author} {\bibfnamefont {V.~S.}\ \bibnamefont
  {Berezinsky}}\ and\ \bibinfo {author} {\bibfnamefont {G.~T.}\ \bibnamefont
  {Zatsepin}},\ }\href@noop {} {\bibfield  {journal} {\bibinfo  {journal} {Sov.
  J. Nucl. Phys.}\ }\textbf {\bibinfo {volume} {11}},\ \bibinfo {pages} {111}
  (\bibinfo {year} {1970})}\BibitemShut {NoStop}%
\bibitem [{\citenamefont {Greisen}(1966)}]{Greisen:1966jv}%
  \BibitemOpen
  \bibfield  {author} {\bibinfo {author} {\bibfnamefont {K.}~\bibnamefont
  {Greisen}},\ }\href {\doibase 10.1103/PhysRevLett.16.748} {\bibfield
  {journal} {\bibinfo  {journal} {Phys. Rev. Lett.}\ }\textbf {\bibinfo
  {volume} {16}},\ \bibinfo {pages} {748} (\bibinfo {year} {1966})}\BibitemShut
  {NoStop}%
\bibitem [{\citenamefont {Zatsepin}\ and\ \citenamefont
  {Kuzmin}(1966)}]{Zatsepin:1966jv}%
  \BibitemOpen
  \bibfield  {author} {\bibinfo {author} {\bibfnamefont {G.~T.}\ \bibnamefont
  {Zatsepin}}\ and\ \bibinfo {author} {\bibfnamefont {V.~A.}\ \bibnamefont
  {Kuzmin}},\ }\href@noop {} {\bibfield  {journal} {\bibinfo  {journal} {JETP
  Lett.}\ }\textbf {\bibinfo {volume} {4}},\ \bibinfo {pages} {78} (\bibinfo
  {year} {1966})}\BibitemShut {NoStop}%
\bibitem [{\citenamefont {Anchordoqui}\ \emph
  {et~al.}(2006{\natexlab{a}})\citenamefont {Anchordoqui}, \citenamefont
  {Cooper-Sarkar}, \citenamefont {Hooper},\ and\ \citenamefont
  {Sarkar}}]{Anchordoqui:2006ta}%
  \BibitemOpen
  \bibfield  {author} {\bibinfo {author} {\bibfnamefont {L.~A.}\ \bibnamefont
  {Anchordoqui}}, \bibinfo {author} {\bibfnamefont {A.~M.}\ \bibnamefont
  {Cooper-Sarkar}}, \bibinfo {author} {\bibfnamefont {D.}~\bibnamefont
  {Hooper}}, \ and\ \bibinfo {author} {\bibfnamefont {S.}~\bibnamefont
  {Sarkar}},\ }\href {\doibase 10.1103/PhysRevD.74.043008} {\bibfield
  {journal} {\bibinfo  {journal} {Phys. Rev.}\ }\textbf {\bibinfo {volume}
  {D74}},\ \bibinfo {pages} {043008} (\bibinfo {year} {2006}{\natexlab{a}})},\
  \Eprint {http://arxiv.org/abs/hep-ph/0605086} {arXiv:hep-ph/0605086}
  \BibitemShut {NoStop}%
\bibitem [{\citenamefont {Martin}\ \emph {et~al.}(2009)\citenamefont {Martin},
  \citenamefont {Stirling}, \citenamefont {Thorne},\ and\ \citenamefont
  {Watt}}]{Martin:2009iq}%
  \BibitemOpen
  \bibfield  {author} {\bibinfo {author} {\bibfnamefont {A.~D.}\ \bibnamefont
  {Martin}}, \bibinfo {author} {\bibfnamefont {W.~J.}\ \bibnamefont
  {Stirling}}, \bibinfo {author} {\bibfnamefont {R.~S.}\ \bibnamefont
  {Thorne}}, \ and\ \bibinfo {author} {\bibfnamefont {G.}~\bibnamefont
  {Watt}},\ }\href {\doibase 10.1140/epjc/s10052-009-1072-5} {\bibfield
  {journal} {\bibinfo  {journal} {Eur. Phys. J.}\ }\textbf {\bibinfo {volume}
  {C63}},\ \bibinfo {pages} {189} (\bibinfo {year} {2009})},\ \Eprint
  {http://arxiv.org/abs/0901.0002} {arXiv:0901.0002 [hep-ph]} \BibitemShut
  {NoStop}%
\bibitem [{\citenamefont {Gandhi}\ \emph {et~al.}(1998)\citenamefont {Gandhi},
  \citenamefont {Quigg}, \citenamefont {Reno},\ and\ \citenamefont
  {Sarcevic}}]{Gandhi:1998ri}%
  \BibitemOpen
  \bibfield  {author} {\bibinfo {author} {\bibfnamefont {R.}~\bibnamefont
  {Gandhi}}, \bibinfo {author} {\bibfnamefont {C.}~\bibnamefont {Quigg}},
  \bibinfo {author} {\bibfnamefont {M.~H.}\ \bibnamefont {Reno}}, \ and\
  \bibinfo {author} {\bibfnamefont {I.}~\bibnamefont {Sarcevic}},\ }\href
  {\doibase 10.1103/PhysRevD.58.093009} {\bibfield  {journal} {\bibinfo
  {journal} {Phys. Rev.}\ }\textbf {\bibinfo {volume} {D58}},\ \bibinfo {pages}
  {093009} (\bibinfo {year} {1998})},\ \Eprint
  {http://arxiv.org/abs/hep-ph/9807264} {arXiv:hep-ph/9807264} \BibitemShut
  {NoStop}%
\bibitem [{\citenamefont {Cooper-Sarkar}\ and\ \citenamefont
  {Sarkar}(2008)}]{CooperSarkar:2007cv}%
  \BibitemOpen
  \bibfield  {author} {\bibinfo {author} {\bibfnamefont {A.}~\bibnamefont
  {Cooper-Sarkar}}\ and\ \bibinfo {author} {\bibfnamefont {S.}~\bibnamefont
  {Sarkar}},\ }\href {\doibase 10.1088/1126-6708/2008/01/075} {\bibfield
  {journal} {\bibinfo  {journal} {JHEP}\ }\textbf {\bibinfo {volume} {01}},\
  \bibinfo {pages} {075} (\bibinfo {year} {2008})},\ \Eprint
  {http://arxiv.org/abs/0710.5303} {arXiv:0710.5303 [hep-ph]} \BibitemShut
  {NoStop}%
\bibitem [{\citenamefont {Jeong}\ and\ \citenamefont
  {Reno}(2010)}]{Jeong:2010za}%
  \BibitemOpen
  \bibfield  {author} {\bibinfo {author} {\bibfnamefont {Y.~S.}\ \bibnamefont
  {Jeong}}\ and\ \bibinfo {author} {\bibfnamefont {M.~H.}\ \bibnamefont
  {Reno}},\ }\href {\doibase 10.1103/PhysRevD.81.114012} {\bibfield  {journal}
  {\bibinfo  {journal} {Phys. Rev.}\ }\textbf {\bibinfo {volume} {D81}},\
  \bibinfo {pages} {114012} (\bibinfo {year} {2010})},\ \Eprint
  {http://arxiv.org/abs/1001.4175} {arXiv:1001.4175 [hep-ph]} \BibitemShut
  {NoStop}%
\bibitem [{\citenamefont {Block}\ \emph {et~al.}(2011)\citenamefont {Block}
  \emph {et~al.}}]{Block:2010fk}%
  \BibitemOpen
  \bibfield  {author} {\bibinfo {author} {\bibfnamefont {M.}~\bibnamefont
  {Block}} \emph {et~al.},\ }\href {\doibase 10.1103/PhysRevD.83.054009}
  {\bibfield  {journal} {\bibinfo  {journal} {Phys. Rev.}\ }\textbf {\bibinfo
  {volume} {D83}},\ \bibinfo {pages} {054009} (\bibinfo {year} {2011})},\
  \Eprint {http://arxiv.org/abs/1010.2486} {arXiv:1010.2486 [hep-ph]}
  \BibitemShut {NoStop}%
\bibitem [{\citenamefont {Martin}\ \emph {et~al.}(2010)\citenamefont {Martin},
  \citenamefont {Stirling}, \citenamefont {Thorne},\ and\ \citenamefont
  {Watt}}]{Martin:2010db}%
  \BibitemOpen
  \bibfield  {author} {\bibinfo {author} {\bibfnamefont {A.~D.}\ \bibnamefont
  {Martin}}, \bibinfo {author} {\bibfnamefont {W.~J.}\ \bibnamefont
  {Stirling}}, \bibinfo {author} {\bibfnamefont {R.~S.}\ \bibnamefont
  {Thorne}}, \ and\ \bibinfo {author} {\bibfnamefont {G.}~\bibnamefont
  {Watt}},\ }\href@noop {} {\bibfield  {journal} {\bibinfo  {journal} {Eur.
  Phys. J.}\ }\textbf {\bibinfo {volume} {C70}},\ \bibinfo {pages} {51}
  (\bibinfo {year} {2010})},\ \Eprint {http://arxiv.org/abs/1007.2624}
  {arXiv:1007.2624 [hep-ph]} \BibitemShut {NoStop}%
\bibitem [{\citenamefont {Aaron}\ \emph {et~al.}(2010)\citenamefont {Aaron}
  \emph {et~al.}}]{:2009wt}%
  \BibitemOpen
  \bibfield  {author} {\bibinfo {author} {\bibfnamefont {F.~D.}\ \bibnamefont
  {Aaron}} \emph {et~al.} (\bibinfo {collaboration} {H1 and ZEUS}),\ }\href
  {\doibase 10.1007/JHEP01(2010)109} {\bibfield  {journal} {\bibinfo  {journal}
  {JHEP}\ }\textbf {\bibinfo {volume} {01}},\ \bibinfo {pages} {109} (\bibinfo
  {year} {2010})},\ \Eprint {http://arxiv.org/abs/0911.0884} {arXiv:0911.0884
  [hep-ex]} \BibitemShut {NoStop}%
\bibitem [{\citenamefont {Goncalves}\ and\ \citenamefont
  {Hepp}(2010)}]{Goncalves:2010ay}%
  \BibitemOpen
  \bibfield  {author} {\bibinfo {author} {\bibfnamefont {V.~P.}\ \bibnamefont
  {Goncalves}}\ and\ \bibinfo {author} {\bibfnamefont {P.}~\bibnamefont
  {Hepp}},\ }\href@noop {} {\  (\bibinfo {year} {2010})},\ \Eprint
  {http://arxiv.org/abs/1011.2718} {arXiv:1011.2718 [hep-ph]} \BibitemShut
  {NoStop}%
\bibitem [{\citenamefont {Thorne}(2005)}]{Thorne:2005kj}%
  \BibitemOpen
  \bibfield  {author} {\bibinfo {author} {\bibfnamefont {R.~S.}\ \bibnamefont
  {Thorne}},\ }\href {\doibase 10.1103/PhysRevD.71.054024} {\bibfield
  {journal} {\bibinfo  {journal} {Phys. Rev.}\ }\textbf {\bibinfo {volume}
  {D71}},\ \bibinfo {pages} {054024} (\bibinfo {year} {2005})},\ \Eprint
  {http://arxiv.org/abs/hep-ph/0501124} {arXiv:hep-ph/0501124} \BibitemShut
  {NoStop}%
\bibitem [{\citenamefont {Froissart}(1961)}]{Froissart:1961ux}%
  \BibitemOpen
  \bibfield  {author} {\bibinfo {author} {\bibfnamefont {M.}~\bibnamefont
  {Froissart}},\ }\href {\doibase 10.1103/PhysRev.123.1053} {\bibfield
  {journal} {\bibinfo  {journal} {Phys. Rev.}\ }\textbf {\bibinfo {volume}
  {123}},\ \bibinfo {pages} {1053} (\bibinfo {year} {1961})}\BibitemShut
  {NoStop}%
\bibitem [{\citenamefont {Berger}\ \emph {et~al.}(2008)\citenamefont {Berger},
  \citenamefont {Block}, \citenamefont {McKay},\ and\ \citenamefont
  {Tan}}]{Berger:2007ic}%
  \BibitemOpen
  \bibfield  {author} {\bibinfo {author} {\bibfnamefont {E.~L.}\ \bibnamefont
  {Berger}}, \bibinfo {author} {\bibfnamefont {M.~M.}\ \bibnamefont {Block}},
  \bibinfo {author} {\bibfnamefont {D.~W.}\ \bibnamefont {McKay}}, \ and\
  \bibinfo {author} {\bibfnamefont {C.-I.}\ \bibnamefont {Tan}},\ }\href
  {\doibase 10.1103/PhysRevD.77.053007} {\bibfield  {journal} {\bibinfo
  {journal} {Phys. Rev.}\ }\textbf {\bibinfo {volume} {D77}},\ \bibinfo {pages}
  {053007} (\bibinfo {year} {2008})},\ \Eprint {http://arxiv.org/abs/0708.1960}
  {arXiv:0708.1960 [hep-ph]} \BibitemShut {NoStop}%
\bibitem [{\citenamefont {Block}\ \emph {et~al.}(2010)\citenamefont {Block},
  \citenamefont {Ha},\ and\ \citenamefont {McKay}}]{Block:2010ud}%
  \BibitemOpen
  \bibfield  {author} {\bibinfo {author} {\bibfnamefont {M.~M.}\ \bibnamefont
  {Block}}, \bibinfo {author} {\bibfnamefont {P.}~\bibnamefont {Ha}}, \ and\
  \bibinfo {author} {\bibfnamefont {D.~W.}\ \bibnamefont {McKay}},\ }\href
  {\doibase 10.1103/PhysRevD.82.077302} {\bibfield  {journal} {\bibinfo
  {journal} {Phys. Rev.}\ }\textbf {\bibinfo {volume} {D82}},\ \bibinfo {pages}
  {077302} (\bibinfo {year} {2010})},\ \Eprint {http://arxiv.org/abs/1008.4555}
  {arXiv:1008.4555 [hep-ph]} \BibitemShut {NoStop}%
\bibitem [{\citenamefont {Amsler}\ \emph {et~al.}(2008)\citenamefont {Amsler}
  \emph {et~al.}}]{Amsler:2008zzb}%
  \BibitemOpen
  \bibfield  {author} {\bibinfo {author} {\bibfnamefont {C.}~\bibnamefont
  {Amsler}} \emph {et~al.} (\bibinfo {collaboration} {Particle Data Group}),\
  }\href {\doibase 10.1016/j.physletb.2008.07.018} {\bibfield  {journal}
  {\bibinfo  {journal} {Phys. Lett.}\ }\textbf {\bibinfo {volume} {B667}},\
  \bibinfo {pages} {1} (\bibinfo {year} {2008})}\BibitemShut {NoStop}%
\bibitem [{\citenamefont {Bulmahn}\ and\ \citenamefont
  {Reno}(2010)}]{Bulmahn:2009ub}%
  \BibitemOpen
  \bibfield  {author} {\bibinfo {author} {\bibfnamefont {A.}~\bibnamefont
  {Bulmahn}}\ and\ \bibinfo {author} {\bibfnamefont {M.~H.}\ \bibnamefont
  {Reno}},\ }\href {\doibase 10.1103/PhysRevD.81.053003} {\bibfield  {journal}
  {\bibinfo  {journal} {Phys. Rev.}\ }\textbf {\bibinfo {volume} {D81}},\
  \bibinfo {pages} {053003} (\bibinfo {year} {2010})},\ \Eprint
  {http://arxiv.org/abs/0912.1385} {arXiv:0912.1385 [hep-ph]} \BibitemShut
  {NoStop}%
\bibitem [{\citenamefont {Nadolsky}\ \emph {et~al.}(2008)\citenamefont
  {Nadolsky} \emph {et~al.}}]{Nadolsky:2008zw}%
  \BibitemOpen
  \bibfield  {author} {\bibinfo {author} {\bibfnamefont {P.~M.}\ \bibnamefont
  {Nadolsky}} \emph {et~al.},\ }\href {\doibase 10.1103/PhysRevD.78.013004}
  {\bibfield  {journal} {\bibinfo  {journal} {Phys. Rev.}\ }\textbf {\bibinfo
  {volume} {D78}},\ \bibinfo {pages} {013004} (\bibinfo {year} {2008})},\
  \Eprint {http://arxiv.org/abs/0802.0007} {arXiv:0802.0007 [hep-ph]}
  \BibitemShut {NoStop}%
\bibitem [{\citenamefont {Connolly}(2006)}]{Connolly:2006gh}%
  \BibitemOpen
  \bibfield  {author} {\bibinfo {author} {\bibfnamefont {A.}~\bibnamefont
  {Connolly}},\ }\href {\doibase 10.1142/S0217751X06033568} {\bibfield
  {journal} {\bibinfo  {journal} {Int. J. Mod. Phys.}\ }\textbf {\bibinfo
  {volume} {A21S1}},\ \bibinfo {pages} {163} (\bibinfo {year}
  {2006})}\BibitemShut {NoStop}%
\bibitem [{\citenamefont {Achterberg}\ \emph {et~al.}(2006)\citenamefont
  {Achterberg} \emph {et~al.}}]{Achterberg:2006md}%
  \BibitemOpen
  \bibfield  {author} {\bibinfo {author} {\bibfnamefont {A.}~\bibnamefont
  {Achterberg}} \emph {et~al.} (\bibinfo {collaboration} {IceCube}),\ }\href
  {\doibase 10.1016/j.astropartphys.2006.06.007} {\bibfield  {journal}
  {\bibinfo  {journal} {Astropart. Phys.}\ }\textbf {\bibinfo {volume} {26}},\
  \bibinfo {pages} {155} (\bibinfo {year} {2006})},\ \Eprint
  {http://arxiv.org/abs/astro-ph/0604450} {arXiv:astro-ph/0604450} \BibitemShut
  {NoStop}%
\bibitem [{\citenamefont {Hoffman}(2010)}]{HoffmanArena}%
  \BibitemOpen
  \bibfield  {author} {\bibinfo {author} {\bibfnamefont {K.}~\bibnamefont
  {Hoffman}},\ }\href@noop {} {\enquote {\bibinfo {title} {{Askaryan Radio
  Array}},}\ } (\bibinfo {year} {2010}),\ \bibinfo {note} {\\ {\tt
  http://indico.in2p3.fr/conferenceOtherViews.py?view=standard\&confId=2719}}\BibitemShut
  {NoStop}%
\bibitem [{\citenamefont {Gerhardt}\ \emph {et~al.}(2010)\citenamefont
  {Gerhardt} \emph {et~al.}}]{Gerhardt:2010js}%
  \BibitemOpen
  \bibfield  {author} {\bibinfo {author} {\bibfnamefont {L.}~\bibnamefont
  {Gerhardt}} \emph {et~al.},\ }\href {\doibase 10.1016/j.nima.2010.09.032}
  {\bibfield  {journal} {\bibinfo  {journal} {Nucl. Instrum. Meth.}\ }\textbf
  {\bibinfo {volume} {A624}},\ \bibinfo {pages} {85} (\bibinfo {year}
  {2010})},\ \Eprint {http://arxiv.org/abs/1005.5193} {arXiv:1005.5193
  [astro-ph.IM]} \BibitemShut {NoStop}%
\bibitem [{\citenamefont {Kusenko}\ and\ \citenamefont
  {Weiler}(2002)}]{Kusenko:2001gj}%
  \BibitemOpen
  \bibfield  {author} {\bibinfo {author} {\bibfnamefont {A.}~\bibnamefont
  {Kusenko}}\ and\ \bibinfo {author} {\bibfnamefont {T.~J.}\ \bibnamefont
  {Weiler}},\ }\href {\doibase 10.1103/PhysRevLett.88.161101} {\bibfield
  {journal} {\bibinfo  {journal} {Phys. Rev. Lett.}\ }\textbf {\bibinfo
  {volume} {88}},\ \bibinfo {pages} {161101} (\bibinfo {year} {2002})},\
  \Eprint {http://arxiv.org/abs/hep-ph/0106071} {arXiv:hep-ph/0106071}
  \BibitemShut {NoStop}%
\bibitem [{\citenamefont {Hooper}(2002)}]{Hooper:2002yq}%
  \BibitemOpen
  \bibfield  {author} {\bibinfo {author} {\bibfnamefont {D.}~\bibnamefont
  {Hooper}},\ }\href {\doibase 10.1103/PhysRevD.65.097303} {\bibfield
  {journal} {\bibinfo  {journal} {Phys. Rev.}\ }\textbf {\bibinfo {volume}
  {D65}},\ \bibinfo {pages} {097303} (\bibinfo {year} {2002})},\ \Eprint
  {http://arxiv.org/abs/hep-ph/0203239} {arXiv:hep-ph/0203239} \BibitemShut
  {NoStop}%
\bibitem [{\citenamefont {Anchordoqui}\ \emph {et~al.}(2010)\citenamefont
  {Anchordoqui} \emph {et~al.}}]{Anchordoqui:2010hq}%
  \BibitemOpen
  \bibfield  {author} {\bibinfo {author} {\bibfnamefont {L.~A.}\ \bibnamefont
  {Anchordoqui}} \emph {et~al.},\ }\href {\doibase 10.1103/PhysRevD.82.043001}
  {\bibfield  {journal} {\bibinfo  {journal} {Phys. Rev.}\ }\textbf {\bibinfo
  {volume} {D82}},\ \bibinfo {pages} {043001} (\bibinfo {year} {2010})},\
  \Eprint {http://arxiv.org/abs/1004.3190} {arXiv:1004.3190 [hep-ph]}
  \BibitemShut {NoStop}%
\bibitem [{\citenamefont {Anchordoqui}\ \emph
  {et~al.}(2006{\natexlab{b}})\citenamefont {Anchordoqui}, \citenamefont {Han},
  \citenamefont {Hooper},\ and\ \citenamefont {Sarkar}}]{Anchordoqui:2005ey}%
  \BibitemOpen
  \bibfield  {author} {\bibinfo {author} {\bibfnamefont {L.}~\bibnamefont
  {Anchordoqui}}, \bibinfo {author} {\bibfnamefont {T.}~\bibnamefont {Han}},
  \bibinfo {author} {\bibfnamefont {D.}~\bibnamefont {Hooper}}, \ and\ \bibinfo
  {author} {\bibfnamefont {S.}~\bibnamefont {Sarkar}},\ }\href {\doibase
  10.1016/j.astropartphys.2005.10.006} {\bibfield  {journal} {\bibinfo
  {journal} {Astropart. Phys.}\ }\textbf {\bibinfo {volume} {25}},\ \bibinfo
  {pages} {14} (\bibinfo {year} {2006}{\natexlab{b}})},\ \Eprint
  {http://arxiv.org/abs/hep-ph/0508312} {arXiv:hep-ph/0508312} \BibitemShut
  {NoStop}%
\bibitem [{\citenamefont {Anchordoqui}\ \emph
  {et~al.}(2006{\natexlab{c}})\citenamefont {Anchordoqui}, \citenamefont
  {Feng},\ and\ \citenamefont {Goldberg}}]{Anchordoqui:2005pn}%
  \BibitemOpen
  \bibfield  {author} {\bibinfo {author} {\bibfnamefont {L.~A.}\ \bibnamefont
  {Anchordoqui}}, \bibinfo {author} {\bibfnamefont {J.~L.}\ \bibnamefont
  {Feng}}, \ and\ \bibinfo {author} {\bibfnamefont {H.}~\bibnamefont
  {Goldberg}},\ }\href {\doibase 10.1103/PhysRevLett.96.021101} {\bibfield
  {journal} {\bibinfo  {journal} {Phys. Rev. Lett.}\ }\textbf {\bibinfo
  {volume} {96}},\ \bibinfo {pages} {021101} (\bibinfo {year}
  {2006}{\natexlab{c}})},\ \Eprint {http://arxiv.org/abs/hep-ph/0504228}
  {arXiv:hep-ph/0504228} \BibitemShut {NoStop}%
\bibitem [{\citenamefont {Borriello}\ \emph {et~al.}(2008)\citenamefont
  {Borriello} \emph {et~al.}}]{Borriello:2007cs}%
  \BibitemOpen
  \bibfield  {author} {\bibinfo {author} {\bibfnamefont {E.}~\bibnamefont
  {Borriello}} \emph {et~al.},\ }\href {\doibase 10.1103/PhysRevD.77.045019}
  {\bibfield  {journal} {\bibinfo  {journal} {Phys. Rev.}\ }\textbf {\bibinfo
  {volume} {D77}},\ \bibinfo {pages} {045019} (\bibinfo {year} {2008})},\
  \Eprint {http://arxiv.org/abs/0711.0152} {arXiv:0711.0152 [astro-ph]}
  \BibitemShut {NoStop}%
\bibitem [{\citenamefont {Romero}\ and\ \citenamefont
  {Sampayo}(2010)}]{Romero:2010zza}%
  \BibitemOpen
  \bibfield  {author} {\bibinfo {author} {\bibfnamefont {I.}~\bibnamefont
  {Romero}}\ and\ \bibinfo {author} {\bibfnamefont {O.~A.}\ \bibnamefont
  {Sampayo}},\ }\href {\doibase 10.1140/epjc/s10052-010-1397-0} {\bibfield
  {journal} {\bibinfo  {journal} {Eur. Phys. J.}\ }\textbf {\bibinfo {volume}
  {C69}},\ \bibinfo {pages} {235} (\bibinfo {year} {2010})}\BibitemShut
  {NoStop}%
\bibitem [{\citenamefont {Hussain}\ \emph {et~al.}(2006)\citenamefont
  {Hussain}, \citenamefont {Marfatia}, \citenamefont {McKay},\ and\
  \citenamefont {Seckel}}]{Hussain:2006wg}%
  \BibitemOpen
  \bibfield  {author} {\bibinfo {author} {\bibfnamefont {S.}~\bibnamefont
  {Hussain}}, \bibinfo {author} {\bibfnamefont {D.}~\bibnamefont {Marfatia}},
  \bibinfo {author} {\bibfnamefont {D.~W.}\ \bibnamefont {McKay}}, \ and\
  \bibinfo {author} {\bibfnamefont {D.}~\bibnamefont {Seckel}},\ }\href
  {\doibase 10.1103/PhysRevLett.97.161101} {\bibfield  {journal} {\bibinfo
  {journal} {Phys. Rev. Lett.}\ }\textbf {\bibinfo {volume} {97}},\ \bibinfo
  {pages} {161101} (\bibinfo {year} {2006})},\ \Eprint
  {http://arxiv.org/abs/hep-ph/0606246} {arXiv:hep-ph/0606246} \BibitemShut
  {NoStop}%
\bibitem [{\citenamefont {Hussain}\ \emph {et~al.}(2008)\citenamefont
  {Hussain}, \citenamefont {Marfatia},\ and\ \citenamefont
  {McKay}}]{Hussain:2007ba}%
  \BibitemOpen
  \bibfield  {author} {\bibinfo {author} {\bibfnamefont {S.}~\bibnamefont
  {Hussain}}, \bibinfo {author} {\bibfnamefont {D.}~\bibnamefont {Marfatia}}, \
  and\ \bibinfo {author} {\bibfnamefont {D.~W.}\ \bibnamefont {McKay}},\ }\href
  {\doibase 10.1103/PhysRevD.77.107304} {\bibfield  {journal} {\bibinfo
  {journal} {Phys. Rev.}\ }\textbf {\bibinfo {volume} {D77}},\ \bibinfo {pages}
  {107304} (\bibinfo {year} {2008})},\ \Eprint {http://arxiv.org/abs/0711.4374}
  {arXiv:0711.4374 [hep-ph]} \BibitemShut {NoStop}%
\bibitem [{\citenamefont {Askaryan}(1962)}]{Askaryan:1962}%
  \BibitemOpen
  \bibfield  {author} {\bibinfo {author} {\bibfnamefont {G.~A.}\ \bibnamefont
  {Askaryan}},\ }\href@noop {} {\bibfield  {journal} {\bibinfo  {journal}
  {JETP}\ }\textbf {\bibinfo {volume} {14}},\ \bibinfo {pages} {441} (\bibinfo
  {year} {1962})}\BibitemShut {NoStop}%
\bibitem [{\citenamefont {Askaryan}(1965)}]{Askaryan:1965}%
  \BibitemOpen
  \bibfield  {author} {\bibinfo {author} {\bibfnamefont {G.~A.}\ \bibnamefont
  {Askaryan}},\ }\href@noop {} {\bibfield  {journal} {\bibinfo  {journal}
  {JETP}\ }\textbf {\bibinfo {volume} {21}},\ \bibinfo {pages} {658} (\bibinfo
  {year} {1965})}\BibitemShut {NoStop}%
\bibitem [{\citenamefont {Alvarez-Muniz}\ and\ \citenamefont
  {Zas}(1997)}]{AlvarezMuniz:1997sh}%
  \BibitemOpen
  \bibfield  {author} {\bibinfo {author} {\bibfnamefont {J.}~\bibnamefont
  {Alvarez-Muniz}}\ and\ \bibinfo {author} {\bibfnamefont {E.}~\bibnamefont
  {Zas}},\ }\href {\doibase 10.1016/S0370-2693(97)01009-5} {\bibfield
  {journal} {\bibinfo  {journal} {Phys. Lett.}\ }\textbf {\bibinfo {volume}
  {B411}},\ \bibinfo {pages} {218} (\bibinfo {year} {1997})},\ \Eprint
  {http://arxiv.org/abs/astro-ph/9706064} {arXiv:astro-ph/9706064} \BibitemShut
  {NoStop}%
\bibitem [{\citenamefont {Saltzberg}\ \emph {et~al.}(2001)\citenamefont
  {Saltzberg} \emph {et~al.}}]{Saltzberg:2000bk}%
  \BibitemOpen
  \bibfield  {author} {\bibinfo {author} {\bibfnamefont {D.}~\bibnamefont
  {Saltzberg}} \emph {et~al.},\ }\href {\doibase 10.1103/PhysRevLett.86.2802}
  {\bibfield  {journal} {\bibinfo  {journal} {Phys. Rev. Lett.}\ }\textbf
  {\bibinfo {volume} {86}},\ \bibinfo {pages} {2802} (\bibinfo {year}
  {2001})},\ \Eprint {http://arxiv.org/abs/hep-ex/0011001}
  {arXiv:hep-ex/0011001} \BibitemShut {NoStop}%
\bibitem [{\citenamefont {Gorham}()}]{gorham}%
  \BibitemOpen
  \bibfield  {author} {\bibinfo {author} {\bibfnamefont {P.}~\bibnamefont
  {Gorham}},\ }\href@noop {} {}\bibinfo {note} {Personal
  communication.}\BibitemShut {Stop}%
\bibitem [{\citenamefont {Gonzalez-Garcia}\ \emph {et~al.}(2008)\citenamefont
  {Gonzalez-Garcia}, \citenamefont {Halzen}, \citenamefont {Maltoni},\ and\
  \citenamefont {Tanaka}}]{GonzalezGarcia:2007gg}%
  \BibitemOpen
  \bibfield  {author} {\bibinfo {author} {\bibfnamefont {M.~C.}\ \bibnamefont
  {Gonzalez-Garcia}}, \bibinfo {author} {\bibfnamefont {F.}~\bibnamefont
  {Halzen}}, \bibinfo {author} {\bibfnamefont {M.}~\bibnamefont {Maltoni}}, \
  and\ \bibinfo {author} {\bibfnamefont {H.~K.~M.}\ \bibnamefont {Tanaka}},\
  }\href {\doibase 10.1103/PhysRevLett.100.061802} {\bibfield  {journal}
  {\bibinfo  {journal} {Phys. Rev. Lett.}\ }\textbf {\bibinfo {volume} {100}},\
  \bibinfo {pages} {061802} (\bibinfo {year} {2008})},\ \Eprint
  {http://arxiv.org/abs/0711.0745} {arXiv:0711.0745 [hep-ph]} \BibitemShut
  {NoStop}%
\bibitem [{\citenamefont {Antoniadis}\ \emph {et~al.}(1998)\citenamefont
  {Antoniadis}, \citenamefont {Arkani-Hamed}, \citenamefont {Dimopoulos},\ and\
  \citenamefont {Dvali}}]{Antoniadis:1998ig}%
  \BibitemOpen
  \bibfield  {author} {\bibinfo {author} {\bibfnamefont {I.}~\bibnamefont
  {Antoniadis}}, \bibinfo {author} {\bibfnamefont {N.}~\bibnamefont
  {Arkani-Hamed}}, \bibinfo {author} {\bibfnamefont {S.}~\bibnamefont
  {Dimopoulos}}, \ and\ \bibinfo {author} {\bibfnamefont {G.~R.}\ \bibnamefont
  {Dvali}},\ }\href {\doibase 10.1016/S0370-2693(98)00860-0} {\bibfield
  {journal} {\bibinfo  {journal} {Phys. Lett.}\ }\textbf {\bibinfo {volume}
  {B436}},\ \bibinfo {pages} {257} (\bibinfo {year} {1998})},\ \Eprint
  {http://arxiv.org/abs/hep-ph/9804398} {arXiv:hep-ph/9804398} \BibitemShut
  {NoStop}%
\bibitem [{\citenamefont {Arkani-Hamed}\ \emph {et~al.}(1998)\citenamefont
  {Arkani-Hamed}, \citenamefont {Dimopoulos},\ and\ \citenamefont
  {Dvali}}]{ArkaniHamed:1998rs}%
  \BibitemOpen
  \bibfield  {author} {\bibinfo {author} {\bibfnamefont {N.}~\bibnamefont
  {Arkani-Hamed}}, \bibinfo {author} {\bibfnamefont {S.}~\bibnamefont
  {Dimopoulos}}, \ and\ \bibinfo {author} {\bibfnamefont {G.~R.}\ \bibnamefont
  {Dvali}},\ }\href {\doibase 10.1016/S0370-2693(98)00466-3} {\bibfield
  {journal} {\bibinfo  {journal} {Phys. Lett.}\ }\textbf {\bibinfo {volume}
  {B429}},\ \bibinfo {pages} {263} (\bibinfo {year} {1998})},\ \Eprint
  {http://arxiv.org/abs/hep-ph/9803315} {arXiv:hep-ph/9803315} \BibitemShut
  {NoStop}%
\bibitem [{\citenamefont {Anchordoqui}\ \emph {et~al.}(2003)\citenamefont
  {Anchordoqui}, \citenamefont {Feng}, \citenamefont {Goldberg},\ and\
  \citenamefont {Shapere}}]{Anchordoqui:2003jr}%
  \BibitemOpen
  \bibfield  {author} {\bibinfo {author} {\bibfnamefont {L.~A.}\ \bibnamefont
  {Anchordoqui}}, \bibinfo {author} {\bibfnamefont {J.~L.}\ \bibnamefont
  {Feng}}, \bibinfo {author} {\bibfnamefont {H.}~\bibnamefont {Goldberg}}, \
  and\ \bibinfo {author} {\bibfnamefont {A.~D.}\ \bibnamefont {Shapere}},\
  }\href {\doibase 10.1103/PhysRevD.68.104025} {\bibfield  {journal} {\bibinfo
  {journal} {Phys. Rev.}\ }\textbf {\bibinfo {volume} {D68}},\ \bibinfo {pages}
  {104025} (\bibinfo {year} {2003})},\ \Eprint
  {http://arxiv.org/abs/hep-ph/0307228} {arXiv:hep-ph/0307228} \BibitemShut
  {NoStop}%
\bibitem [{\citenamefont {Alvarez-Muniz}\ \emph {et~al.}(2002)\citenamefont
  {Alvarez-Muniz}, \citenamefont {Feng}, \citenamefont {Halzen}, \citenamefont
  {Han},\ and\ \citenamefont {Hooper}}]{AlvarezMuniz:2002ga}%
  \BibitemOpen
  \bibfield  {author} {\bibinfo {author} {\bibfnamefont {J.}~\bibnamefont
  {Alvarez-Muniz}}, \bibinfo {author} {\bibfnamefont {J.~L.}\ \bibnamefont
  {Feng}}, \bibinfo {author} {\bibfnamefont {F.}~\bibnamefont {Halzen}},
  \bibinfo {author} {\bibfnamefont {T.}~\bibnamefont {Han}}, \ and\ \bibinfo
  {author} {\bibfnamefont {D.}~\bibnamefont {Hooper}},\ }\href {\doibase
  10.1103/PhysRevD.65.124015} {\bibfield  {journal} {\bibinfo  {journal} {Phys.
  Rev.}\ }\textbf {\bibinfo {volume} {D65}},\ \bibinfo {pages} {124015}
  (\bibinfo {year} {2002})},\ \Eprint {http://arxiv.org/abs/hep-ph/0202081}
  {arXiv:hep-ph/0202081} \BibitemShut {NoStop}%
\bibitem [{\citenamefont {Aaltonen}\ \emph {et~al.}(2008)\citenamefont
  {Aaltonen} \emph {et~al.}}]{Aaltonen:2008hh}%
  \BibitemOpen
  \bibfield  {author} {\bibinfo {author} {\bibfnamefont {T.}~\bibnamefont
  {Aaltonen}} \emph {et~al.} (\bibinfo {collaboration} {CDF}),\ }\href
  {\doibase 10.1103/PhysRevLett.101.181602} {\bibfield  {journal} {\bibinfo
  {journal} {Phys. Rev. Lett.}\ }\textbf {\bibinfo {volume} {101}},\ \bibinfo
  {pages} {181602} (\bibinfo {year} {2008})},\ \Eprint
  {http://arxiv.org/abs/0807.3132} {arXiv:0807.3132 [hep-ex]} \BibitemShut
  {NoStop}%
\bibitem [{\citenamefont {Abazov}\ \emph {et~al.}(2009)\citenamefont {Abazov}
  \emph {et~al.}}]{:2009mh}%
  \BibitemOpen
  \bibfield  {author} {\bibinfo {author} {\bibfnamefont {V.~M.}\ \bibnamefont
  {Abazov}} \emph {et~al.} (\bibinfo {collaboration} {D0}),\ }\href {\doibase
  10.1103/PhysRevLett.103.191803} {\bibfield  {journal} {\bibinfo  {journal}
  {Phys. Rev. Lett.}\ }\textbf {\bibinfo {volume} {103}},\ \bibinfo {pages}
  {191803} (\bibinfo {year} {2009})},\ \Eprint {http://arxiv.org/abs/0906.4819}
  {arXiv:0906.4819 [hep-ex]} \BibitemShut {NoStop}%
\bibitem [{\citenamefont {Khachatryan}\ \emph {et~al.}(2011)\citenamefont
  {Khachatryan} \emph {et~al.}}]{Khachatryan:2010wx}%
  \BibitemOpen
  \bibfield  {author} {\bibinfo {author} {\bibfnamefont {V.}~\bibnamefont
  {Khachatryan}} \emph {et~al.} (\bibinfo {collaboration} {CMS}),\ }\href
  {\doibase 10.1016/j.physletb.2011.02.032} {\bibfield  {journal} {\bibinfo
  {journal} {Phys. Lett.}\ }\textbf {\bibinfo {volume} {B697}},\ \bibinfo
  {pages} {434} (\bibinfo {year} {2011})},\ \Eprint
  {http://arxiv.org/abs/1012.3375} {arXiv:1012.3375 [hep-ex]} \BibitemShut
  {NoStop}%
\bibitem [{\citenamefont {Yoshida}(2010)}]{Yoshida:2010kp}%
  \BibitemOpen
  \bibfield  {author} {\bibinfo {author} {\bibfnamefont {S.}~\bibnamefont
  {Yoshida}},\ }\href {\doibase 10.1103/PhysRevD.82.103012} {\bibfield
  {journal} {\bibinfo  {journal} {Phys. Rev.}\ }\textbf {\bibinfo {volume}
  {D82}},\ \bibinfo {pages} {103012} (\bibinfo {year} {2010})},\ \Eprint
  {http://arxiv.org/abs/1009.1679} {arXiv:1009.1679 [hep-ph]} \BibitemShut
  {NoStop}%
\bibitem [{\citenamefont {Engel}\ \emph {et~al.}(2001)\citenamefont {Engel},
  \citenamefont {Seckel},\ and\ \citenamefont {Stanev}}]{Engel:2001hd}%
  \BibitemOpen
  \bibfield  {author} {\bibinfo {author} {\bibfnamefont {R.}~\bibnamefont
  {Engel}}, \bibinfo {author} {\bibfnamefont {D.}~\bibnamefont {Seckel}}, \
  and\ \bibinfo {author} {\bibfnamefont {T.}~\bibnamefont {Stanev}},\ }\href
  {\doibase 10.1103/PhysRevD.64.093010} {\bibfield  {journal} {\bibinfo
  {journal} {Phys. Rev.}\ }\textbf {\bibinfo {volume} {D64}},\ \bibinfo {pages}
  {093010} (\bibinfo {year} {2001})},\ \bibinfo {note} {we take the muon flavor
  component from Figure 9 and then extract the electron flavor component by
  using the electron to muon ratios derived from Figure 4},\ \Eprint
  {http://arxiv.org/abs/astro-ph/0101216} {arXiv:astro-ph/0101216} \BibitemShut
  {NoStop}%
\bibitem [{\citenamefont {Casal~Larana}()}]{CasalLarana:2010zz}%
  \BibitemOpen
  \bibfield  {author} {\bibinfo {author} {\bibfnamefont {B.}~\bibnamefont
  {Casal~Larana}},\ }\href@noop {} {\ }\bibinfo {note}
  {FERMILAB-THESIS-2010-04}\BibitemShut {NoStop}%
\bibitem [{\citenamefont {Neyman}\ and\ \citenamefont
  {Pearson}(1933)}]{NeymanPearson}%
  \BibitemOpen
  \bibfield  {author} {\bibinfo {author} {\bibfnamefont {J.}~\bibnamefont
  {Neyman}}\ and\ \bibinfo {author} {\bibfnamefont {E.}~\bibnamefont
  {Pearson}},\ }\href@noop {} {\bibfield  {journal} {\bibinfo  {journal} {Phil.
  Trans. of the Royal Soc. of London}\ }\textbf {\bibinfo {volume} {A31}},\
  \bibinfo {pages} {289} (\bibinfo {year} {1933})}\BibitemShut {NoStop}%
\end{thebibliography}
\end{document}